\documentclass[aps,prx,showpacs,floatfix,twocolumn,superscriptaddress]{revtex4-1}
\usepackage{graphicx}
\usepackage{bm,amsmath}
\usepackage{dcolumn}
\usepackage{amsfonts,amssymb}
\usepackage{chemfig}
\begin{document}

\title{Nearly deconfined spinon excitations \\ in the square-lattice spin-$1/2$  Heisenberg antiferromagnet}

\author{Hui Shao}
\email{shaohui@csrc.ac.cn}
\affiliation{Beijing Computational Science Research Center, Beijing 100193, China}
\affiliation{Department of Physics, Boston University, 590 Commonwealth Avenue, Boston, Massachusetts 02215, USA}

\author{Yan Qi Qin}
\affiliation{Institute of Physics, Chinese Academy of Sciences, Beijing 100190, China}
\affiliation{School of Physical Sciences, University of Chinese Academy of Sciences, Beijing 100190, China}

\author{Sylvain Capponi}
\affiliation{Laboratoire de Physique Théorique, Université de Toulouse and CNRS, UPS (IRSAMC), F-31062 Toulouse, France}
\affiliation{Department of Physics, Boston University, 590 Commonwealth Avenue, Boston, Massachusetts 02215, USA}

\author{Stefano Chesi}
\affiliation{Beijing Computational Science Research Center, Beijing 100193, China}
\author{Zi Yang Meng}

\email{zymeng@iphy.ac.cn}
\affiliation{Institute of Physics, Chinese Academy of Sciences, Beijing 100190, China}
\affiliation{School of Physical Sciences, University of Chinese Academy of Sciences, Beijing 100190, China}

\author{Anders W. Sandvik}
\email{sandvik@bu.edu}
\affiliation{Department of Physics, Boston University, 590 Commonwealth Avenue, Boston, Massachusetts 02215, USA}
\affiliation{Beijing Computational Science Research Center, Beijing 100193, China}
\date{\today}

\begin{abstract}
  We study the spin excitation spectrum (dynamic structure factor) of the spin-$1/2$ square-lattice Heisenberg antiferromagnet and
  an extended model (the $J$-$Q$ model) including four-spin interactions $Q$ in addition to the Heisenberg exchange $J$. Using an improved
  method for stochastic analytic continuation of imaginary-time correlation functions computed with quantum Monte Carlo simulations, we
  can treat the sharp ($\delta$-function) contribution to the structure factor expected from spinwave (magnon) excitations, in addition to
  resolving a continuum above the magnon energy. Spectra for the Heisenberg model are in excellent agreement with recent neutron scattering
  experiments on Cu(DCOO)$_2$$\cdot$4D$_2$O, where a broad spectral-weight continuum at wavevector $q=(\pi,0)$ was interpreted as deconfined
  spinons, i.e., fractional excitations carrying half of the spin of a magnon. Our results at $(\pi,0)$ show a similar reduction of the magnon
  weight and a large continuum, while the continuum is much smaller at $q=(\pi/2,\pi/2)$ (as also seen experimentally). We further
  investigate the reasons for the small magnon weight at $(\pi,0)$ and the nature of the corresponding excitation by studying the evolution of
  the spectral functions in the $J$-$Q$ model. Upon turning on the $Q$ interaction, we observe a rapid reduction of the magnon weight to zero, well before
  the system undergoes a deconfined quantum phase transition into a non-magnetic spontaneously dimerized state. Based on these results, we re-interpret
  the picture of deconfined spinons at $(\pi,0)$ in the experiments as nearly deconfined spinons---a precursor to deconfined quantum criticality.
  To further elucidate the picture of a fragile $(\pi,0)$-magnon pole in the Heisenberg model and its depletion in the $J$-$Q$ model, we introduce an
  effective model of the excitations in which a magnon can split into two spinons which do not separate but fluctuate in and out of the magnon
  space (in analogy with the resonance between a photon and a particle-hole pair in the exciton-polariton problem). The model can reproduce the
  reduction of magnon weight and lowered excitation energy at $(\pi,0)$ in the Heisenberg model, as well as the energy maximum and smaller continuum
  at $(\pi/2,\pi/2)$. It can also account for the rapid loss of the $(\pi,0)$ magnon with increasing $Q$ and a remarkable persistence of a
  large magnon pole at $q=(\pi/2,\pi/2)$ even at the deconfined critical point. The fragility of the magnons close to $(\pi,0)$ in the Heisenberg
  model suggests that various interactions that likely are important in many materials, e.g., longer-range pair exchange, ring exchange, and
  spin-phonon interactions, may also destroy these magnons and lead to even stronger spinon signatures than in Cu(DCOO)$_2$$\cdot$4D$_2$O.
\end{abstract}

\maketitle

\section{Introduction}

The spin $S=1/2$ antiferromagnetic (AFM) Heisenberg model is the natural starting point for describing the magnetic properties of many electronic
insulators with localized spins \cite{Anderson52}. The two-dimensional (2D) square-lattice variant of the model came to particular prominence due to its
relevance to the undoped parent compounds of the cuprate high-temperature superconductors \cite{Anderson87,Manousakis91}, e.g., \chemfig{La_2CuO_4},
and it has also remained a fruitful testing grounds for quantum magnetism more broadly. Though there is no rigorous proof of the existence of
AFM long-range order at temperature $T=0$ in the case of $S=1/2$ spins (while for $S \ge 1$ there is such a proof \cite{Neves86}), series-expansion
\cite{Singh89} and quantum Monte Carlo (QMC) calculations \cite{Reger88,Runge92,sandvik97,sandvik10a,Jiang11} have convincingly demonstrated a sublattice
magnetization in close agreement with the simple linear spinwave theory. Thermodynamic properties and the spin correlations at $T>0$ \cite{Beard96,Kim98,Beard98}
also conform very nicely with the expectations \cite{Chakravarty89,Hasenfratz93} for a ``renormalized classical'' system with exponentially divergent correlation
length when $T \to 0$. Thus, at first sight it may appear that the case is settled and the system lacks 'exotic' quantum-mechanical features. However, it has
been known for some time that the dynamical properties of the model at short wavelengths cannot be fully described by spinwave theory. Along the line
${\bf q}=(\pi,0)$ to $(\pi/2,\pi/2)$ in the  Brillouin zone (BZ) of the square lattice (with lattice spacing one), the magnon energy is maximal and constant
within linear spinwave theory. However, various numerical calculations have pointed to a significant suppression of the magnon energy and an anomalously large
continuum of excitations in the dynamic spin structure factor $S({\bf q},\omega)$ around ${\bf q}=(\pi,0)$ \cite{Singh95,sandvik01,zheng05,Powalski15,Powalski17}.
At ${\bf q}=(\pi/2,\pi/2)$ the energy is instead elevated and the continuum is smaller. Conventional spinwave theory can only capture a small fraction 
of the $(\pi,0)$
anomaly, even when pushed to high orders in the $1/S$ expansion \cite{Igarashi92,Canali93,Igarashi05,syromyatnikov10}.

A large continuum at high energies for ${\bf q}$ close to $(\pi,0)$ was also observed in neutron scattering experiments on \chemfig{La_2CuO_4}, but an opposite
trend in the energy shifts is apparent there; a reduction at ${\bf q}=(\pi/2,\pi/2)$ and increase at $(\pi,0)$ \cite{coldea01,headings10}. It was realized that
this is due to the fact that the exchange constant $J$ is large in this case ($J\approx 100 {\rm meV}$), and, when considering its origin from an electronic Hubbard
model, higher-order exchange processes play an important role \cite{Peres02,Wan09,Delannoy09,Piazza12}. Interestingly, in \chemfig{Cu{(DCOO)}_2\cdot4D_2O} (CFTD), which is
considered the best realization of the square-lattice Heisenberg model to date, anomalous features in close agreement with those in the Heisenberg model have been
observed \cite{ronnow01,christensen07,piazza15}. In this case the exchange constant is much smaller, $J\approx 6 {\rm meV}$, and the higher-order interactions
are expected to be relatively much smaller than in \chemfig{La_2CuO_4}.

The existence of a large continuum in the excitation spectrum close to ${\bf q}=(\pi,0)$ has for some time prompted speculations of physics beyond magnons in
materials such as \chemfig{La_2CuO_4} and CFTD. In particular, in recent low-temperature polarized neutron scattering experiments on CFTD \cite{piazza15}, the
broad and spin-isotropic continuum in $S({\bf q},\omega)$ at ${\bf q}=(\pi,0)$ was interpreted as a sign of deconfinement of spinons, i.e., that the $S=1$
degrees of freedom excited by a neutron at this wavevector would fractionalize into two independently propagating $S=1/2$ objects. In contrast, the
$(\pi/2,\pi/2)$ scattering remained more magnon-like, with a small spin-anisotropic continuum. Calculation within a class of variational resonating-valence-bond 
(RVB) wave functions gave some support to this picture \cite{piazza15}, showing that a pair of spinons originating from a ``broken''
valence bond \cite{tang13} at ${\bf q}=(\pi,0)$ could deconfine and account for both the energy suppression and the broad continuum.

A potential problem with the spinon interpretation is that there is still also a magnon pole at ${\bf q}=(\pi,0)$, even though its amplitude is suppressed, and
this would indicate that the lowest-energy excitations there are still magnons. Lacking AFM long-range order, the RVB wave-function does not contain
any magnon pole, and the interplay between the magnon and putative spinon continuum was not considered in Ref.~\onlinecite{piazza15}. Many
different calculations have indicated a magnon pole in the entire BZ in 2D Heisenberg model \cite{Singh95,sandvik01,zheng05,Powalski15,Powalski17}.
The prominent continuum at and close to ${\bf q}=(\pi,0)$ has been ascribed to multi-magnon processes, and systematic expansions \cite{Powalski17} in the
number of magnons indeed converge rapidly and give results for the relative weight of the single-magnon pole in close agreement \cite{Powalski17note} with
series-expansion and QMC calculations \cite{Singh95,sandvik01}. Since the results also agree very well with the neutron data for CFTD, the spinon
interpretation of the experiments can be questioned.

Despite the apparent success of the multi-magnon scenario in accounting for the observations, one may still wonder whether spinons could
have some relevance in the Heisenberg model and materials such as CFTD and \chemfig{La_2CuO_4}---this question is the topic of the present paper. Our main motivation
for revisiting the spinon scenario is the direct connection between the Heisenberg model and deconfined quantum criticality: If a certain four-spin interaction
$Q$ is added to the Heisenberg exchange $J$ on the square lattice (the $J$-$Q$ model \cite{sandvik07}), the system can be driven into a spontaneously dimerized
ground state; a valence-bond solid (VBS). At the dimerization point, $Q_c/J \approx 22$, the AFM order also vanishes, in what appears to be a continuous quantum phase transition
\cite{melko08,harada13,shao16}, in accord with the scenario of deconfined quantum critical points \cite{senthil04a,senthil04b}. At the critical point, linearly
dispersing gapless triplets emerge at ${\bf q}=(\pi,0)$ and $(0,\pi)$ \cite{spanu06,suwa16} in addition to the gapless points $(0,0)$ and $(\pi,\pi)$ in the long-range
ordered AFM, and all the low-energy $S=1$ excitations around these points should comprise spinon pairs. Thus, it is possible that the reduction in $(\pi,0)$ excitation
energy observed in the Heisenberg model and CFTD is a precursor to deconfined quantum criticality. If that is the case, then it may indeed be possible to also
describe the continuum in $S(q,\omega)$ around $q=(\pi,0)$ in terms of spinons, as already proposed in Ref.~\onlinecite{Powalski17}. However, the persistence of
the magnon pole remains unexplained in this scenario.

Here we will revise and complete the picture of deconfined spinon states in the continuum by also investigating the nature of the sharp
magnon-like state in the Heisenberg model and its fate as the deconfined critical point is approached. Using QMC calculations and an improved numerical analytic
continuation technique (also presented in this paper) to obtain the dynamic structure factor from imaginary-time dependent spin correlations, we will show
that the $(\pi,0)$ magnon pole in the Heisenberg model is fragile---it is destroyed in the presence of even a very small $Q$ interaction, well before the
critical point where the AFM order vanishes. In contrast, the $(\pi/2,\pi/2)$ magnon is robust and survives even at the critical point. We will explain
these behaviors within an effective magnon-spinon mixing model, where a bare magnon in the Heisenberg model becomes dressed by fluctuating in and out of a
two-spinon continuum at higher energy. The mixing is the strongest at ${\bf q}=(\pi,0)$; the point of minimum gap between the magnon and spinon. Our results indicate that there
already exist spinons close above the magnon band in the Heisenberg model, and a small perturbation, here the $Q$ interaction, can cause their bare energy
to dip below the magnon, thus destabilizing this part of the magnon band and changing the nature of the excitation from a well-defined magnon-spinon resonance to
a broad continuum of spinon states. In contrast, the $(\pi/2,\pi/2)$ spinons, which are at their dispersion maximum, never fall below the magnon energy,
thus explaining the robust magnon in this case.

The proximity of the square-lattice Heisenberg AFM to a so-called AF* phase has been proposed as the reason for the $(\pi,0)$ anomaly \cite{piazza15}.
The AF* phase has topological $Z_2$ order but still also has AFM long-range order, and it hosts gapped spinon excitations in addition to low-energy magnons
\cite{balents99,senthil00}. In our scenario it is instead the proximity to a VBS and the intervening deconfined quantum critical point that is responsible for the
presence of high-energy spinons and the excitation anomaly in the Heisenberg model. Our results for the $J$-$Q$ model show that the ${\bf q}=(\pi,0)$ magnon pole
is very fragile in the Heisenberg model and the magnon picture should fail completely around this wavevector even with a rather weak deformation of the
model, likely also with other perturbations than the $Q$-term considered here (e.g., frustrated further-neighbor couplings, ring exchange, or perhaps
even spin-phonon couplings). Thus, although the almost ideal Heisenberg magnet CFTD should only host nearly deconfined spinons, other materials may possibly
have sufficient additional quantum fluctuations to cause full deconfinement close to ${\bf q}=(\pi,0)$.

Our numerical results for $S(q,\omega)$ rely heavily on an improved stochastic method for analytic continuation of QMC-computed imaginary-time
correlation functions. It allows us to test for the presence of a $\delta$-function in the spectral function and determine its weight. In Sec.~\ref{sec:sac}
we will summarize the features of the method that are of critical importance to the present work (leaving more extensive discussions of a broader range of
applications of similar ideas for a future publication \cite{shao17}). We also present tests using synthetic data, which show that the kind of spectral
function expected in the Heisenberg model indeed can be reproduced with QMC data of typical quality. Readers who are not interested in technical details
can skip this section and go directly to Sec.~\ref{sec:hberg}, where we present a brief recapitulation of the key aspects of the method before discussing
the dynamic structure factor of the Heisenberg model. In addition to the QMC results, we also compare with Lanczos exact diagonalization (ED) results for
small systems and study finite-size behaviors with both methods. We compare our results with the recent experimental data for CFTD. In Sec.~\ref{sec:jq}
we discuss results for the $J$-$Q$ model, focusing on the points ${\bf q}=(\pi,0)$ and ${\bf q}=(\pi/2,\pi/2)$, where the excitation spectrum evolves in
completely different ways as the $Q$-interactions are increased and the deconfined critical point is approached. In Sec.~\ref{sec:heff} we present
the effective magnon-spinon mixing model for the excitations and discuss numerical solutions of it. We summarize and further discuss our main
conclusions in Sec.~\ref{sec:summary}.

\section{Stochastic Analytic Continuation}\label{sec:sac}

We will consider a spectral function---the dynamic spin structure factor---at temperature $T=0$. A general spectral function of any bosonic
operator $O$ can be written in the basis of eigenstates $|n\rangle$ and eigenvalues $E_n$ of the Hamiltonian as
\begin{equation}
S(\omega)=\frac{1}{\pi} \sum_{n}|\langle n|O|0\rangle|^2\delta(\omega - [E_n-E_0]).
\label{somegasum}
\end{equation}
For the dynamic spin structure factor $S({\bf q},\omega)$ at momentum transfer ${\bf q}$ and energy transfer $\omega$, the corresponding
operator is the Fourier transform of a spin operator, e.g., the $z$ component
\begin{equation}
O^z_q = \frac{1}{\sqrt{N}}\sum_{i=1}^N {\rm e}^{-i {\bf r}_i \cdot {\bf q}} S^z_i,
\label{szft}
\end{equation}
where ${\bf r}_i$ is the coordinate of site $i$; here on the square lattice with the lattice spacing set to unity.
In this section we will keep the discussion general and do not need to consider the form of the operator.

\subsection{Preliminaries} 
\label{sec:sac1}

In QMC simulations we calculate the corresponding correlation function in imaginary time,
\begin{equation}
G(\tau)=\langle O(\tau)O(0)\rangle,
\label{otau}
\end{equation}
where $O(\tau) = {\rm e}^{\tau H} O {\rm e}^{-\tau H}$, and its relationship to the real-frequency spectral function is
\begin{equation}
G(\tau) = \int_{0}^\infty d\omega S(\omega){\rm e}^{-\tau \omega}.
\label{gtau}
\end{equation}
Some QMC methods, such as the SSE method \cite{sse1} applied here to the Heisenberg model, can provide an unbiased
stochastic approximation $\bar G_i \equiv\bar G(\tau_i)$ to the true correlation function $G_i\equiv G(\tau_i)$ for a set of imaginary
times $\tau_i$, $i=1,\ldots,N_\tau$ \cite{sse2,sse3}. These data points have statistical errors $\sigma_i$ (one standard
deviation of the mean value). Since the statistical errors are correlated, their full characterization requires the
covariance matrix, which can be evaluated with the QMC data divided up into a large number of bins. Denoting the QMC bin averages
by $G^{b}_{i}$ for bins $b=1,2,\ldots,N_B$, we have $\bar G_i = \sum_b G^{b}_{i}/N_B$ and the covariance matrix is given by
\begin{equation}
  C_{ij} = \frac{1}{N_B(N_B-1)}\sum_{b=1}^{N_B} (G^b_i-\bar G_i)(G^b_j-\bar G_j),
  \label{covmat}
\end{equation}
where we also assume that the bins are based on sufficiently long simulations to be statistically independent.
The diagonal elements of $C$ are the squares of the standard statistical errors; $\sigma_i^2 = C_{ii}$.

In a numerical analytic continuation procedure, the spectral function is parametrized in some way, e.g., with a large number of
$\delta$-functions on a dense grid of frequencies or with adjustable positions in the frequency continuum. The parameters (e.g.,
the amplitudes of the $\delta$-functions) are adjusted for compatibility with the QMC data using the relationship Eq.~(\ref{gtau}).
Given a proposal for $S(\omega)$, there is then a set of numbers $\{G_i\}$ whose closeness to the corresponding
QMC-computed function is quantified in the standard way in a data-fitting procedure by the ``goodness of the fit''
\begin{equation}
\chi^2 = \sum_{i=1}^{N_\tau}\sum_{j=1}^{N_\tau} (G_i-\bar G_i)C^{-1}_{ij}(G_j-\bar G_j).
\label{chi2}
\end{equation}
In practice, we compute the eigenvalues $\epsilon_i$ and eigenvectors of $C$ and transform the kernel ${\rm e}^{-\tau \omega}$ of 
Eq.~(\ref{gtau}) to this basis. With $\Delta_i=G_i-\bar G_i$ transformed to the same basis, the
goodness of the fit is diagonal;
\begin{equation}
\chi^2 = \sum_{i=1}^{N_\tau} \left ( \frac{\Delta_i}{\epsilon_i} \right )^2,
\label{chi2dia}
\end{equation}
and can be more rapidly evaluated.

A reliable diagonalization of the covariance matrix requires more than $N_\tau$ bins and we here typically use at least $10 \times N_\tau$ bins,
with $N_\tau$ in the range $50-100$ and the $\tau$ points chosen on a uniform or quadratic grid. We evaluate the covariance matrix (\ref{covmat})
by bootstrapping, with the total number of bootstrap samples (each consisting of $N_B$ random selections out of the $N_B$ bins) even
larger than the number of bins. In Appendix \ref{app:covar} we show some examples of covariance eigenvalues and eigenvectors.

Minimizing $\chi^2$ does not produce useful results. If positive-definiteness of the spectrum is imposed, the ``best''
solution consists of a typically small number of sharp peaks \cite{schuttler86,sandvik98}, and there are many other very different solutions with almost
the same $\chi^2$-value, reflecting the ill-posed nature of the inverse of the Laplace transform in Eq.~(\ref{gtau}). Without positive-definiteness 
the problem is even more ill-posed. Some regularization mechanism therefore has to be applied.

In the standard Maximum-Entropy (ME) method \cite{gull84,silver90,jarrell96}, an entropy $E$,
\begin{equation}
E = - \int_0^\infty d\omega S(\omega)\ln \left ( \frac{S(\omega)}{D(\omega)} \right ),
\label{entropy}
\end{equation}
of the spectrum with respect to a ``default model'' $D(\omega)$ is defined (i.e., $E$ is maximized when $S=D$), and the data is taken into
account by maximizing the function
\begin{equation}
Q=\alpha E-\chi^2.
\label{qalpha}
\end{equation}
This produces the most likely spectrum, given the data and the entropic prior. Different variants of the
method prescribe different ways of determining the parameter $\alpha$, or, in some variants, results are averaged over $\alpha$.

Here we will use stochastic analytic continuation~\cite{sandvik98,beach04,syljuasen08,fuchs10} (SAC), where the entropy is not imposed explicitly
as a prior but is generated implicitly by a Monte Carlo sampling procedure of a suitably parametrized spectrum. We will introduce a parametrization that
enables us to study a spectrum containing a sharp $\delta$-function, which is impossible to resolve with the standard ME approaches (and also with 
standard SAC) because of the low entropy of such spectra.

\subsection{Sampling Procedures}

Following one of the main lines of the SAC approach \cite{sandvik98,beach04,syljuasen08,fuchs10}, we sample the spectrum with a
probability distribution resembling the Boltzmann distribution of a statistical-mechanics problem, with $\chi^2/2$ playing the role of the energy of
a system at a fictitious temperature $\Theta$;
\begin{equation}
P(S) \propto \exp \left (-\frac{\chi^2}{2\Theta} \right ).
\label{pofs}
\end{equation}
Lowering $\Theta$ leads to less fluctuations and a smaller mean value $\langle \chi^2\rangle$, and this parameter therefore plays a regularization
role similar to $\alpha$ in the ME function, Eq.~(\ref{qalpha}) \cite{beach04}. Several proposals for how to choose the value of
$\Theta$ have been put forward \cite{sandvik98,beach04,syljuasen08,fuchs10}. There is also another line of SAC methods in which good spectra
(in the sense of low $\chi^2$ values) are generated not by sampling at a fictitious temperature, but according to some other distribution with
other regularizing parameters \cite{mishchenko02}. Using Eq.~(\ref{pofs}) allows us to construct direct analogues with statistical mechanics,
e.g., as concerns configurational entropy \cite{sandvik16}. Before describing our scheme of fixing $\Theta$, we discuss a parametrization of the
spectrum specifically adapted to the dynamic spin structure factor of interest in this work.

\begin{figure}[t]
\centering
\includegraphics[width=80mm]{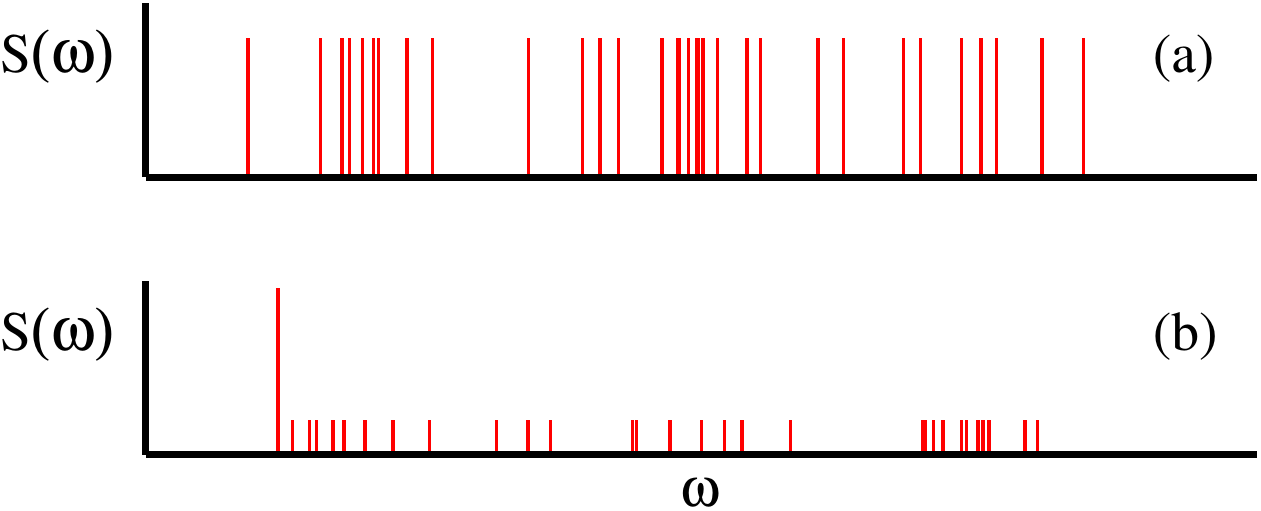}
\caption{Parametrizations of the spectral function used in this work. In (a) a large number of $\delta$-functions with the same amplitude occupy
  frequencies $\omega_i$ in the continuum (or, in practice, on a very fine frequency grid). The locations are sampled in the SAC procedure.
  In (b), the $\delta$-function at the lowest frequency $\omega_0$ has a larger amplitude, $a_0 > a_i$ for $i>0$, and this amplitude is
  optimized in the way described in the text. The frequencies of all the $\delta$-functions, including $\omega_0$, are sampled as in (a),
  but with the constraint $\omega_0 < \omega_i ~\forall ~i>0$.}
\label{deltas}
\end{figure}

We parametrize the spectrum by a number $N_{\omega}$ of $\delta$-functions in the continuum, as illustrated
in Fig.~\ref{deltas};
\begin{equation}
S(\omega) = \sum_{i=0}^{N_\omega-1}{a_i}\delta(\omega-\omega_i),
\label{somegadeltas}
\end{equation}
working with a normalized spectrum, so that
\begin{equation}
\sum_{i=0}^{N_\omega-1}{a_i}=1,
\label{normdeltas}
\end{equation}
which corresponds to $G(0)=1$ in Eq.~(\ref{gtau}). The pre-normalized value of $G(0)$ is used as a factor in the final result.
In sampling the spectrum, we never change the normalization, and $G(0)$ therefore is not included in the data set
defining $\chi^2$ in Eq.~(\ref{chi2dia}). The covariance matrix, Eq.~(\ref{covmat}), is also computed with normalization to $G(0)=1$ for each
bootstrap sample, which has a consequence that the individual statistical errors $\sigma_i \to 0$ for $\tau_i \to 0$, as discussed 
further in Appendix \ref{app:covar}.

In Fig.~\ref{deltas}(a) the $\delta$-functions all have the same weight, $a_i=N^{-1}_\omega$, with $N_\omega$ typically ranging from
$500$ to $2000$ in the calculations presented in this paper. The sampling corresponds to changing the locations (frequencies) $\omega_i$ of
the $\delta$-functions, with the standard Metropolis probability used to accept or reject a change $\omega_i \to \omega_i + d$,
with $d$ chosen at random within a window centered at $d=0$. The width of the window is adjusted to give an acceptance rate close to $1/2$.
We collect the spectral weight in a histogram, averaging over sufficiently many updating cycles of the frequencies to obtain smooth results.
In practice, in order to be able to use a precomputed kernel ${\rm e}^{-\omega_j\tau_i}$ in Eq.~(\ref{gtau}) for all times $\tau_i$ and frequencies
$\omega_j$, we use a very fine grid of allowed frequencies (much finer than the histogram used for collecting the spectrum), e.g., with spacing
$\Delta_\omega=10^{-5}$ in typical cases where the dominant spectral weight is roughly within the range $0-5$. We then also need to
impose a maximum frequency, e.g., $\omega_{\rm max}=20$ under the above conditions. With $\approx 100$ $\tau$-values the amount of memory needed to
store the kernel is then still reasonable, and in practice the fine grid produces results indistinguishable from ones obtained in the
continuum (strictly speaking double-precision floating-point numbers) without limitation, i.e., without even an upper bound imposed on
the frequencies.

We have found that not changing the amplitudes of the $\delta$-functions is an advantage in terms of the sampling time required to obtain good
results, and there are other advantages as well, as will be discussed further in a forthcoming technical article \cite{shao17}. One can also initialize
the amplitudes with a range of different weights (e.g., of the form $a_i \propto i^\alpha$, with $\alpha>0$), while maintaining the
normalization Eq.~(\ref{normdeltas}). This modification of the scheme can help if the spectrum has a gap separating regions of significant
spectral weight, since an additional amplitude-swap update, $a_i \leftrightarrow a_j$, can easily transfer weight between two separate regions
when the weights are all different, thus speeding up the sampling (but we typically do not find significant differences in the final results
as compared with all-equal $a_i$). This method was already applied to spectral functions of a 3D quantum critical antiferromagnet in
Ref.~\onlinecite{qin17}. Here we do not have any indications of mid-spectrum gaps and use the constant-weight ensemble, however, with
a crucial modification.

As illustrated in Fig.~\ref{deltas}(b), in order to reproduce the kind of spectral function expected in the 2D Heisenberg model---a magnon pole
followed by a continuum---we have developed a modified parametrization where we give special treatment to the $\delta$-function with lowest frequency
$\omega_0$. We adjust its amplitude $a_0$ in a manner described further below but keep it fixed in the sampling of frequencies. The common amplitude for
the other $\delta$-functions is then $a_i = (1-a_0)/(N_\omega-1)$. The determination of the best $a_0$ value also relies on how the sampling temperature
$\Theta$ is chosen, which we discuss next.

Consider first the case of all $\delta$-functions having equal amplitude; Fig.~\ref{deltas}(a). As an initial step, we carry out a simulated
annealing procedure with slowly decreasing $\Theta$ to find the lowest, or very close to the lowest, possible value of $\chi^2$ (which will never be
exactly $0$, no matter how many $\delta$-functions are used, because of the positive-definiteness imposed on the spectrum). We then raise $\Theta$ to a
value where the sampled mean value $\langle \chi^2\rangle$ of the goodness of fit is higher than the minimum value $\chi^2_{\rm min}$ by an amount of the
order of the standard deviation of the $\chi^2$ distribution, i.e, going away from the overfitting region where the process becomes sensitive
to the detrimental effects of the statistical errors (i.e., producing a nonphysical spectrum with a small number of sharp peaks). Considering the
statistical expectation that the best fit should have $\chi^2_{\rm min} \approx N_{\rm dof}=N_\tau-N_{\rm para}$, where $N_{\rm para}$ is the (unknown)
effective number of parameters of the spectrum and the minimum $\chi^2$ value can be taken as an estimate of the effective number of degrees of
freedom; $\chi^2_{\rm min} \approx N_{\rm dof}$. Hence, the standard deviation $\sigma_{\chi^2}=(2N_{\rm dof})^{1/2}$ can be replaced by the
statistically valid approximation
\begin{equation}
\sigma_{\chi^2} \approx \sqrt{2\chi^2_{\rm min}}.
\end{equation}
Thus, we adjust $\Theta$ such that
\begin{equation}
\langle \chi^2\rangle \approx \chi^2_{\rm min} +a\sqrt{2\chi^2_{\rm min}}, 
\label{chi2crit}
\end{equation}
with the constant $a$ of order one. For spectral functions with no sharp features, we find that this method with the parametrization in Fig.~\ref{deltas}(a)
produces good, stable results, with very little dependence of the average spectrum on $a$ as long as it is of order one. For $a \to 0$ the data become
overfitted, leading eventually to a spectrum consisting of a small number of sharp peaks with little resemblance to the true spectrum.

Using the unrestricted sampling with the parametrization in Fig.~\ref{deltas}(a), with QMC data of typical quality one cannot expect to resolve
a very sharp peak---in the extreme case a $\delta$-function---because it will be washed out by entropy. Therefore, in most of the calculations reported
in this paper we proceed in a different way in order to incorporate the expected $\delta$-function. After determining $\chi^2_{\rm min}$, we switch to the
parametrization in Fig.~\ref{deltas}(b), and the next step is to find an optimal value of the amplitude $a_0$. To this end we rely on the insight from
Ref.~\onlinecite{sandvik16}
that the optimal value of a parameter affecting the amount of configurational entropy in the spectrum can be determined by monitoring $\langle\chi^2\rangle$
as a function of that parameter at fixed sampling temperature $\Theta$. In the case of $a_0$, increasing its value will remove entropy from the spectrum.
Since entropy is what tends to spread out the spectral weight excessively into regions where there should be little weight or no weight at all, a reduced
entropy can be reflected in a smaller value of $\langle \chi^2\rangle$. Thus, in cases where the spectrum is gapped, a sampling with the parametrization
in Fig.~\ref{deltas}(a) will lead to spectral weight in the gap and an overall distorted spectrum. However, upon switching to the parametrization in
Fig.~\ref{deltas}(b) and gradually increasing $a_0$, no weight can appear below $\omega_0$ and $\langle \omega_0\rangle$ will gradually increase (and
note again that $\omega_0$ is not fixed but is sampled along with the other frequencies $\omega_i$) because a good match with the QMC data $\{\bar G\}$
cannot be obtained if there is too much weight in the gap. In this process $\langle \chi^2\rangle$ will decrease. Upon increasing $a_0$ further,
$\langle \omega_0\rangle$ will eventually be pushed too far above the gap, and then $\langle \chi^2\rangle$ clearly must start to increase. Thus, if there is
a $\delta$-function at the lower edge of the spectrum pursued, one can in general expect a minimum in $\langle \chi^2\rangle$ versus $a_0$, and, if the QMC data
are good enough, this minimum should be close to the true value of $a_0$. When fixing $a_0$ to its optimal value at the $\langle \chi^2\rangle$-minimum, the
frequency $\omega_0$ should fluctuate around its correct value (with normally very small fluctuations so that the final result is a very sharp peak). If
there is no such $\delta$-function in the true spectrum, one would expect the $\langle \chi^2\rangle$ minimum very close to $a_0=0$. Extensive testing,
to be reported elsewhere \cite{shao17}, has confirmed this picture. We here show test results relevant to the type of spectral function expected for
the 2D Heisenberg model.

One might think that we could also sample the weight $a_0$ instead of optimizing its fixed value. The reason why this does not work is at the heart 
of our approach: Including Monte Carlo updates changing the value of $a_0$ (and thus also of all other weights $a_{i>0}$ to maintain normalization), entropic pressures 
will favor values close to the other amplitudes and the results (which we have confirmed) are indistinguishable from those obtained without special 
treatment of the lower edge, i.e., the parametrization in Fig.~\ref{deltas}(a). The entropy associated with different parametrizations will be further 
discussed in a separate article \cite{shao17}.

\subsection{Tests on synthetic data}

To test whether the method can resolve the kind of spectral features that are expected in the 2D Heisenberg model, we construct a synthetic spectral
function with a $\delta$-function of weight $a_0$ and frequency $\omega_0$, followed by a continuum with total weight $1-a_0$. The relationship in
Eq.~(\ref{gtau}) is used to obtain $G(\tau)$ for a set of $\tau$-points and normal-distributed noise is added to the $G$ values, with standard deviation
typical in QMC results. To provide an even closer approximation to real QMC data, we construct correlated noise. Here one can adjust the autocorrelation
time of the correlated noise to be close to what is observed in QMC data. The way we do this is discussed in more detail in Appendix \ref{app:covar}.

As we will discuss in Sec.~\ref{sec:hberg}, for the 2D Heisenberg model we find that the smallest relative weight of the magnon pole is $\approx 0.4$ 
at ${\bf q}=(\pi,0)$. We therefore
here test with $a_0=0.4$, set $\omega_0=1$ and take for the continuum a truncated Gaussian (with no weight below $\omega_0$) of width
$\sigma=1$. This situation of no gap between the $\delta$-function and the continuum should be expected to be very challenging for any
analytic continuation method. Extracting $a_0$ and $\omega_0$ by simply fitting an exponential $a_0{\rm e}^{-\omega_0\tau}$ to the QMC
data for large $\tau$ is difficult because there will never be any purely exponential decay (unlike the case where there is a gap
between the $\delta$-function and the continuum) and the best one could hope for is to extrapolate the parameters based on different
ranges of $\tau$ included in the fit, or with some more sophisticated analysis \cite{suwa16}. As we will see below, with noise levels in the
synthetic data similar to our real QMC data, the SAC procedure outlined above not only produces good results for $a_0$ and $\omega_0$
but also reproduces the continuum well.

\begin{figure}[t]
\centering
\includegraphics[width=75mm, clip]{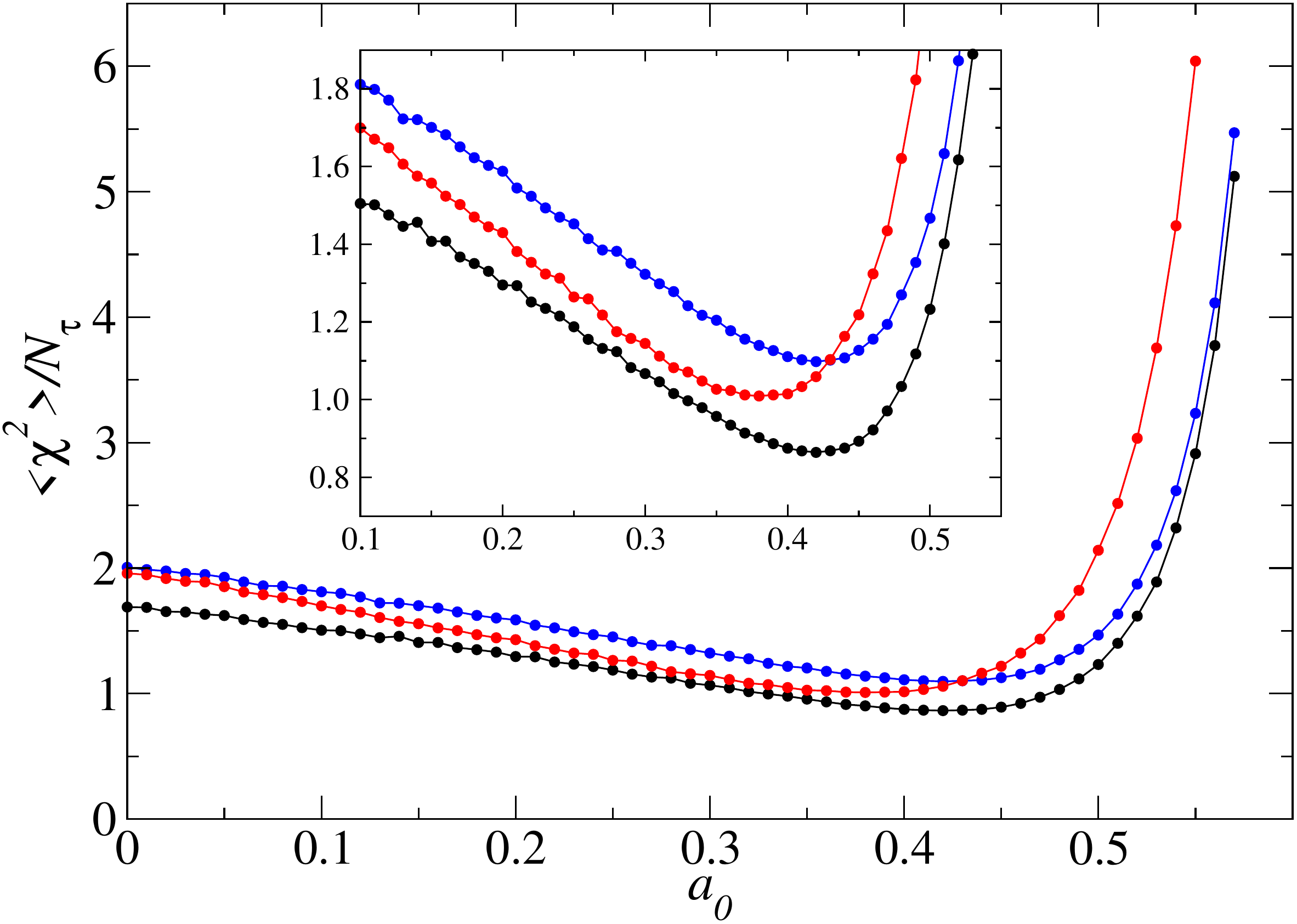}
\caption{The goodness of the fit versus the amplitude $a_0$ of the lowest $\delta$-function in three runs with different
  noise realizations for a synthetic spectrum with a $\delta$-function of weight $a_0=0.4$ at $\omega_0=1$. The continuum is
  a Gaussian of width $1$ centered at the same $\omega_0$, with the weight below $\omega_0$ excluded. The noise level 
  is $\sigma_i \approx 10^{-5}$ and the errors are correlated with autocorrelation time $1$ according to the description
  in Appendix \ref{app:covar}. The inset shows the data close to the $\langle \chi^2\rangle$ minimum on a different scale.}
\label{chi2fig}
\end{figure}

When looking for the minimum value of $\langle \chi^2\rangle$ versus $a_0$, it is better to start with a somewhat higher $\Theta$ than
what is obtained with the $\chi^2$ criterion in Eq.~(\ref{chi2crit}), so that the minimum can be more pronounced. Staying in the regime where the fit
can still be considered good and the effects on $S(\omega)$ of a slightly elevated $\Theta$ are very minor, 
we aim for $\langle \chi^2\rangle \approx \chi^2_{\rm min} + bN_\tau$ with
$b=1$ or $2$ at the initial stage of fixing $\Theta$ without the special treatment of the lowest $\delta$-function. With the so obtained $\Theta$ we scan
over $a_0$ with some step size $\Delta a_0$. The scan is terminated when $\langle \chi^2\rangle$ has increased well past its minimum.
The $\langle \chi^2\rangle$ curve can be analyzed later to locate the optimal $a_0$ value. If all the spectra generated in the  scan have been
saved one can simply use the best one. Since $\langle \chi^2\rangle$ normally will be significantly smaller at the optimal value of $a_0$ than
at the starting point with $a_0=0$, there is typically no need for further adjustments of $\Theta$ later, though one can also do a final run
at the optimal $a_0$ with the criterion in Eq.~(\ref{chi2crit}).

Fig.~\ref{chi2fig} shows typical $\langle \chi^2\rangle$ behaviors in tests with spectrum consisting of a $\delta$-function and a continuum of relative size
and width similar to what we will report for the Heisenberg model in the next section. Here we used $80$ $\tau$-points on a uniform grid with spacing $\Delta_\tau=0.1$
and noise level $\sigma_i \approx 10^{-5}$ for $\tau$ points sufficiently away from $\tau=0$. We built in covariance similar to what is observed in the QMC data
(also discussed in Appendix \ref{app:covar}). We can indeed observe a clear minimum in the $\langle \chi^2\rangle$ curve close to the expected value $a_0=0.4$.
The deviations from this point reflect the effects of the statistical errors. In several runs at much smaller noise level, $\sigma_i \approx 10^{-6}$, the
minimum was always at $0.40$ in scans with $\Delta a_0=0.01$.

\begin{figure}[t]
\centering
\includegraphics[width=70mm, clip]{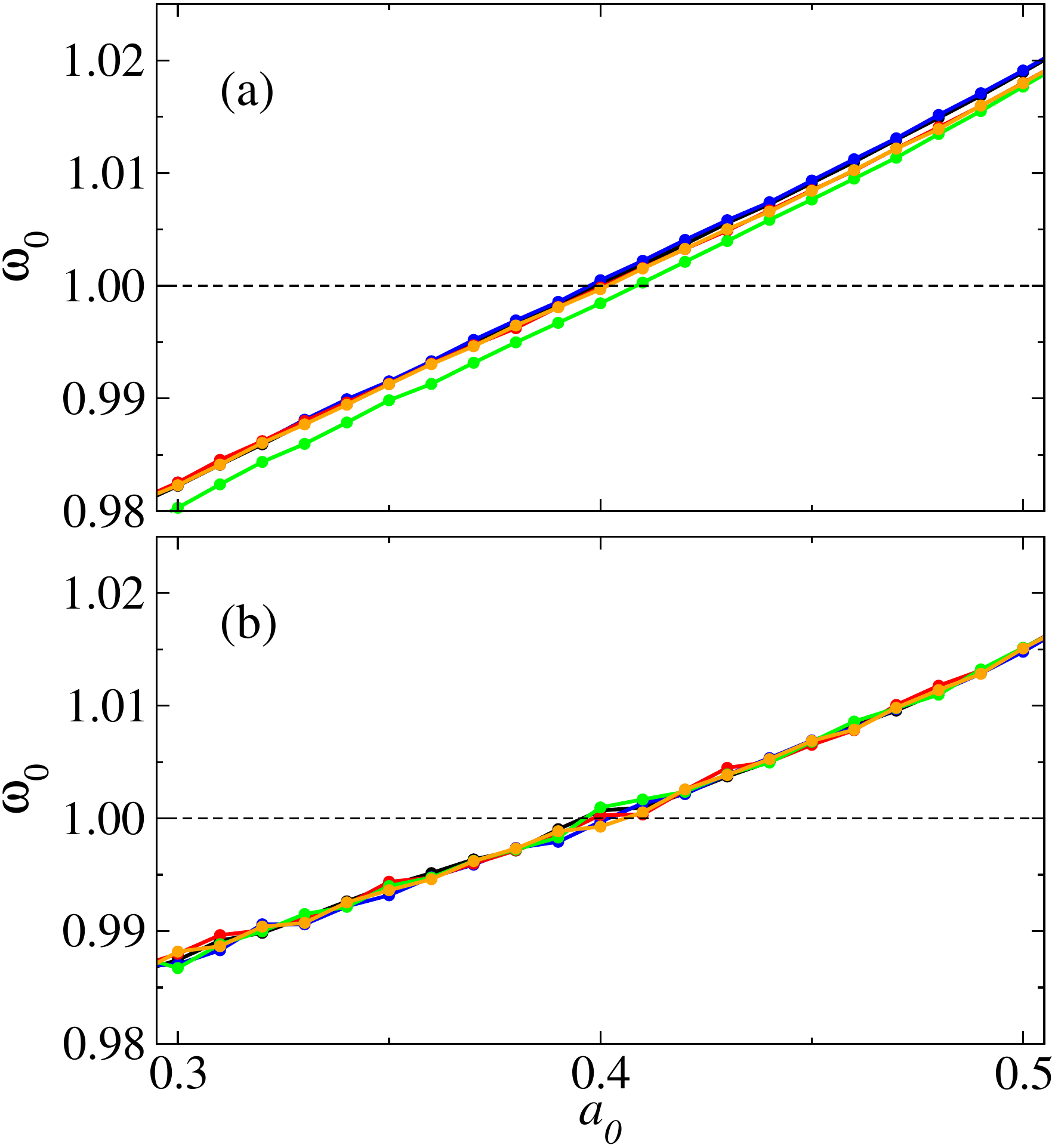}
\caption{Mean value of $\omega_0$ in SAC runs with four different noise realizations (shown in different colors) graphed vs
the amplitude parameter $a_0$. The noise level is $10^{-5}$ and $10^{-6}$ in (a) and (b), respectively, and the number of $\delta$-functions was
$N_\omega=1000$ and $N_\omega=2000$. In both cases $G(\tau)$ values were generated on a uniform grid with $\Delta_\tau=0.1$
for $\tau$ up to the point where the relative error exceeds $10\%$.}
\label{w0}
\end{figure}

The effects of the noise are smaller in the mean location $\langle \omega_0\rangle$ of the lowest $\delta$-function. Fig.~\ref{w0} shows
results versus $a_0$ from several different runs. At the correct value $a_0=0.4$, the error in the frequency is typically less than
$10^{-3}$ at noise level $10^{-5}$ and smaller still at $10^{-6}$. Considering the uncertainty in the location of the minimum
in Fig.~\ref{chi2fig}, the total error on $\omega_0$ of course becomes higher, but still the precision is typically better than $10^{-2}$ for
noise level $10^{-5}$ and much better at $10^{-6}$.

\begin{figure}[t]
\centering
\includegraphics[width=70mm, clip]{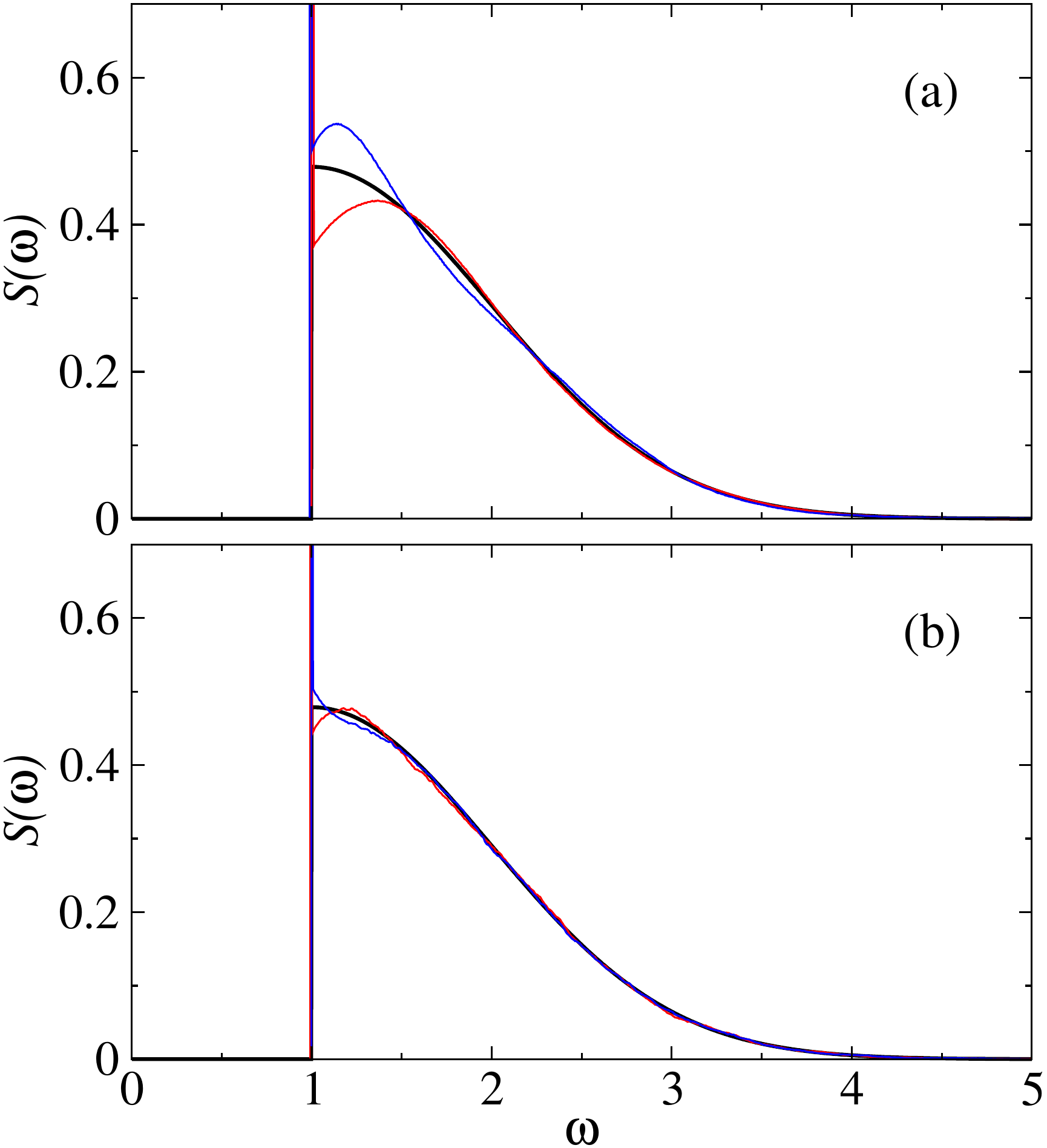}
\caption{Two typical SAC-computed spectral functions (red and blue curves, obtained with different noise realizations) compared with the underlying
  true synthetic spectrum (thicker black curve, with the half-Gaussian containing $60\%$ of the weight). The parameters of the spectrum
  are the same as in Fig.~\ref{chi2}. The noise level is $10^{-5}$ and $10^{-6}$ in (a) and (b), respectively.}
\label{sw}
\end{figure}

The full SAC spectral functions at both noise levels are shown in Fig.~\ref{sw}, for two noise realizations in each case
(with the spectra taken at their respective optimal $a_0$ values). When constructing the histogram for averaging the spectrum, here with a
bin with $\Delta\omega=0.005$, we also include the main $\delta$-peak. If the fluctuations in $\omega_0$ are large, a broadened peak will
result. Here the fluctuations are very small and no significant broadening is seen beyond that due to the histogram binning. As discussed
above, the location of the main peak is very well reproduced. The continuum typically shows the strongest deviations from the correct
curve close to the edge. The improvements when going from noise level $10^{-5}$ to $10^{-6}$ are obvious in the figure.

Statistical errors of order $10^{-5}$ in the correlation function $G(\tau)$ normalized to $1$ at $\tau=0$ are relatively easy to achieve
in QMC calculations, and in many cases it is possible to go to $10^{-6}$ or even better. The tests here show that quite detailed information
can be obtained with such data for spectral functions with a prominent $\delta$-function at the lower edge followed by a broad continuum.
Importantly, the approach also involves the estimation of the statistical error on the weight of the $\delta$-function through a bootstrapping
procedure, and based on tests such as those above, as well as additional cases, we do not see any signs of further systematical errors in the
weight and location of the $\delta$-function, i.e., the method is unbiased in this regard. It is still of course not easy to discriminate
between a spectrum with an extremely narrow peak and one with a true $\delta$-function, but a broad peak will manifest itself in the loss of 
amplitude $a_0$, accumulation of the ``background'' $\delta$-functions as a leading maximum at the edge, and in large fluctuations in the
lower edge $\omega_0$. We therefore have good reasons to believe that the approach is suitable in general both for reproducing spectra with
an extremely narrow peak and for detecting when such a peak is absent.

\section{Heisenberg model}\label{sec:hberg}

In quantum magnetism the most important spectral function is the dynamic spin structure factor $S^\alpha ({\bf q},\omega)$, corresponding to
the correlations of the spin operator $S_{\bf q}^\alpha$ $(\alpha=x,y,z)$, the Fourier transform of the real-space spin operator $S_{\bf r}^\alpha$
as in Eq.~(\ref{szft}). This spectral function is directly proportional to the inelastic neutron-scattering cross-section at wavevector
transfer ${\bf q}$ and energy transfer $\omega$ \cite{formfactor}. In this paper we focus on isotropic spin systems and do not break the symmetry
in the finite-size calculations; thus all components $\alpha$ are the same, corresponding to the total cross-section averaged over the longitudinal
and transverse channels (i.e., as obtained in experiments with unpolarized neutrons). We consider the $z$-component in the SSE-QMC calculations and 
hereafter use the notation $S({\bf q},\omega)$ without any
$\alpha$ superscript. With sufficiently large inverse temperature, here $\beta=4L$ in most QMC simulations, we obtain ground-state properties for all
practical purposes for ${\bf q}$ at which the gap $\omega_{\bf q}$ is sufficiently large. More precisely, we have well-converged data for all
${\bf q}$ except for ${\bf q}=(\pi,\pi)$, where the finite-size gap closes as $1/L^2$ (this being the lowest excitation in the Anderson
tower of quantum-rotor states), much faster than the lowest magnon excitation which has a gap $\propto 1/L$. Therefore, in the following we do not
analyze the not fully-converged ${\bf q}=(\pi,\pi)$ data. In addition to the QMC calculations, where we go up to linear system sizes $L=48$,
we also report exact $T=0$ Lanczos ED results for lattices with up to $N=40$ spins.

For the square-lattice Heisenberg antiferromagnet, the spectral function in calculations such as conventional spinwave expansions
\cite{Igarashi92,Canali93,Igarashi05,syromyatnikov10} and continuous unitary transformations (an approach which also starts from spinwave theory, 
formulated with the Dyson-Maleev representation of the spin operators) \cite{Powalski15,Powalski17} contains a dominant $\delta$-function at 
the lowest frequency $\omega_{\bf q}$ and a continuum above this frequency,
\begin{equation}
S({\bf q},\omega)=S_0({\bf q})\delta(\omega-\omega_{\bf q})+S_c({\bf q},\omega),
\label{Sfunction}
\end{equation}
where $\omega_{\bf q}$ is also the single-magnon dispersion and $S_0({\bf q})$ is the spectral weight in the magnon pole.
We define the relative weight of the single-magnon contribution as
\begin{equation}
a_0({\bf q})=\frac{S_0({\bf q})}{\int d\omega S({\bf q},\omega)},
\label{define-a0}
\end{equation}
in the same way as the generic $a_0$ in Sec.~\ref{sec:sac}.

In principle the single-magnon pole may be broadened, but the damping
processes causing this are of very high order in the spinwave interaction terms and we are not
aware of any calculations estimating these effects quantitatively. In general it is expected that the broadening of the magnon pole itself should be very small
in bipartite (collinear AFM-ordered) Heisenberg systems \cite{chernyshev06,chernyshev09}. Accordingly, we can here make the simplifying assumption that there
is no broadening at $T=0$ of the single-magnon pole itself, i.e., that interaction effects are manifested as spectral weight transferred from the
$\delta$-function to the continuum above it. In contrast, in non-bipartite (frustrated) antiferromagnets with non-collinear order, there are
other lower-order magnon damping mechanisms present that cause significant broadening of the $\delta$-function \cite{chernyshev06,chernyshev09}.

In a previous QMC calculation where the analytic continuation was carried out by function fitting including a $\delta$-function edge \cite{sandvik01},
the continuum $S_c({\bf q},\omega)$ was modeled with a specific functional form with a number of parameters (adjusted to fit the QMC data).
Here we do not make any prior assumptions on the shape of the continuum, instead applying the SAC procedure with the parametrization illustrated
in Fig.~\ref{deltas}(b). If the $\delta$-function is actually substantially broadened, such that the separation of the spectrum into two distinct parts in
Eq.~(\ref{Sfunction}) becomes inappropriate, we expect our SAC approach to simply give a very small amplitude $S_0({\bf q})$ when this is the case.
We will see examples of this kind of full depletion of the magnon pole later in Sec.~\ref{sec:jq}, where other interactions are added to the Heisenberg
model (the $J$-$Q$ model). Later in this section we will also show some results for the Heisenberg model obtained without assuming a $\delta$-function
in Eq.~(\ref{Sfunction}).

To briefly recapitulate the version of SAC we developed in Sec.~\ref{sec:sac}, after fixing a proper sampling temperature using the spectrum without special 
treatment of the leading $\Delta$-function, i.e., the parametrization of the dynamic structure factor illustrated in in Fig.~\ref{deltas}(a), in the final stage 
of the sampling process we use the parametrization ofFig.~\ref{deltas}(b). The amplitude of the leading $\delta$-function is optimized based on the entropic 
signal---a minimum in the mean goodness of the fit, $\langle \chi^2\rangle$. The location of this special $\delta$-function is sampled along with all the other 
``small'' ones representing the continuum, and the spectral weight as a function of the frequency is collected in a histogram (here typically with bin size 
$\Delta\omega=0.005$). Thus,
in the final averaged spectrum the magnon pole may be broadened by fluctuations in its location, but, as we will see below, the width is typically very narrow
and for all practical purposes it remains a $\delta$-function contribution. Here the level of the statistical QMC errors, with the definitions discussed in
Sec.~\ref{sec:sac}, is  $10^{-5}$ or better (some raw data are shown in Appendix \ref{app:covar}). Extensive testing, exemplified in Fig.~\ref{sw}, demonstrates
that the method is well capable of reproducing the type of spectral function of interest here to a good degree with this data quality. The number
$N_\omega$ of $\delta$-functions required in the continuum in order to obtain well converged results depends on the quality of the QMC data. We have
carried out tests with different $N_\omega$ and find good convergence of the results when $N_\omega \approx 500-1000$. The results presented
below were obtained with $N_\omega=2000$.

\subsection{Spectral functions at different wavevectors}

\begin{figure}[t]
\centering
\includegraphics[width=83mm]{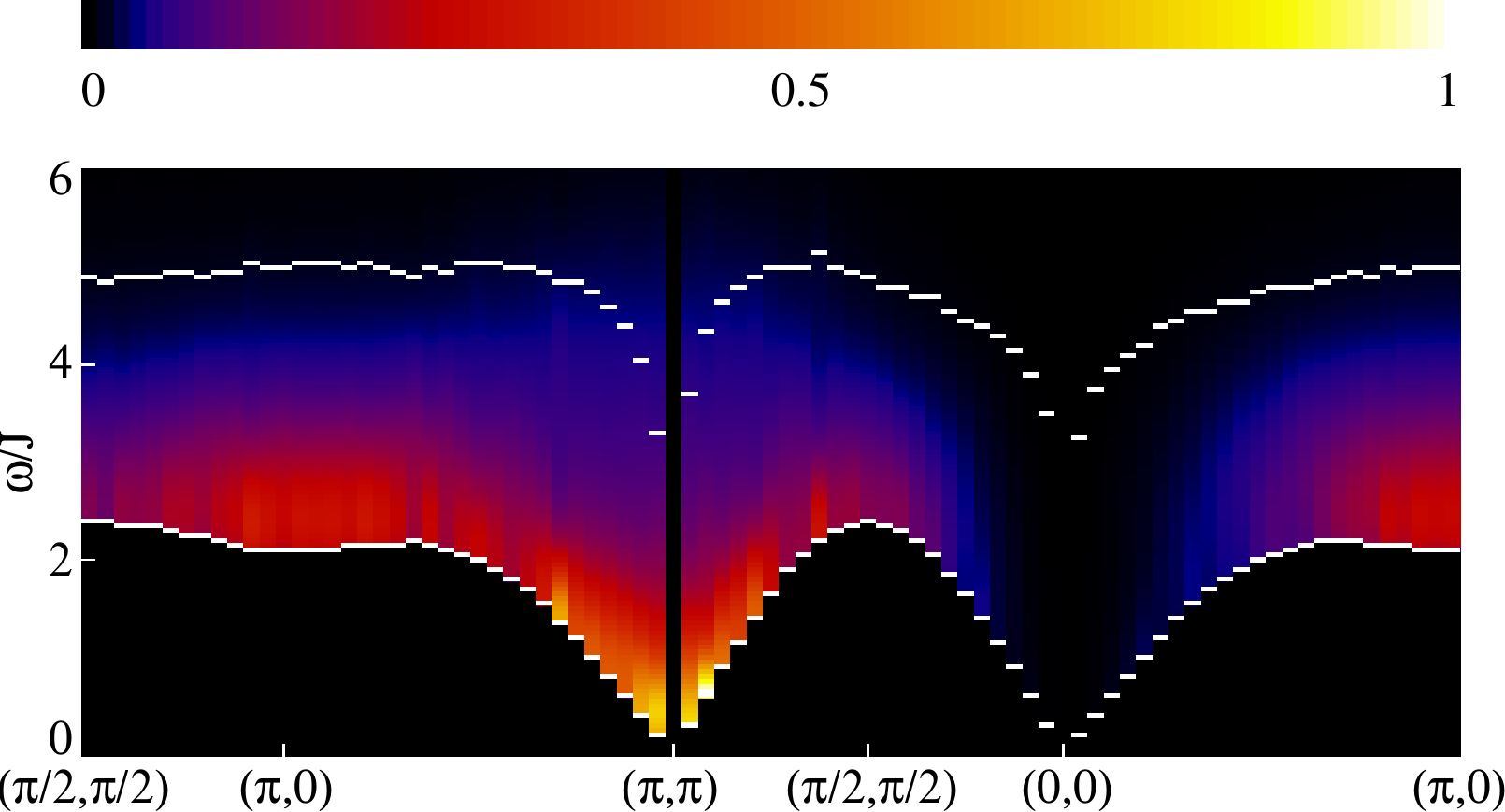}
\caption{The dynamic structure factor of the 2D Heisenberg model computed on an $L=48$ lattice along
the path in the BZ indicated on the $x$-axis. The $y$-axis is the energy transfer $\omega$ in units of the coupling $J$.
The magnon peak ($\delta$-function) at the lower edge of the spectrum is marked in white irrespective of its weight, while
the continuum is shown with color coding on an arbitrary scale where the highest value is $1$. The upper white curve
corresponds to the location where, for given ${\bf q}$, $5\%$ of the spectral weight remains above it.}
\label{haf-1}
\end{figure}

For an overview, we first show the spectral function for the $L=48$ system with a color plot in Fig.~\ref{haf-1}, where the $x$-axis
corresponds to the wavevector along a standard path in the BZ and the $y$-axis is the frequency $\omega$. The location of the magnon
pole (the dispersion relation) is indicated, and for the continuum a color coding is used. We also show an upper spectral bound defined
such that $95\%$ of the weight for each ${\bf q}$ falls between the two curves. Due to matrix-elements effects related to conservation of
the magnetization ($S^z_{q=0}$) of the Heisenberg model, the total spectral weight vanishes as $q\to 0$ and it is seen in Fig.~\ref{haf-1} to
be small in a wide region around this point. Both the total weight and the low-energy scattering is maximized as ${\bf q} \to (\pi,\pi)$. As
mentioned above, exactly at $(\pi,\pi)$ our calculations are not $T \to 0$ converged, and we therefore do not show any results for this case.
The width in $\omega$ of the region in which
$95\%$ of the weight is concentrated is seen to be almost independent on ${\bf q}$. However, since the total spectral weight for ${\bf q}$ close
to $(\pi,\pi)$ is very large there is significant weight extending up to $\omega \approx 6$, while in other ${\bf q}$ regions the weight extends
roughly up to $4.5-5$ [except close to $(0,0)$, where no significant weight can be discerned in the density plot with the color coding used].

\begin{figure}[t]
\centering
\includegraphics[width=65mm]{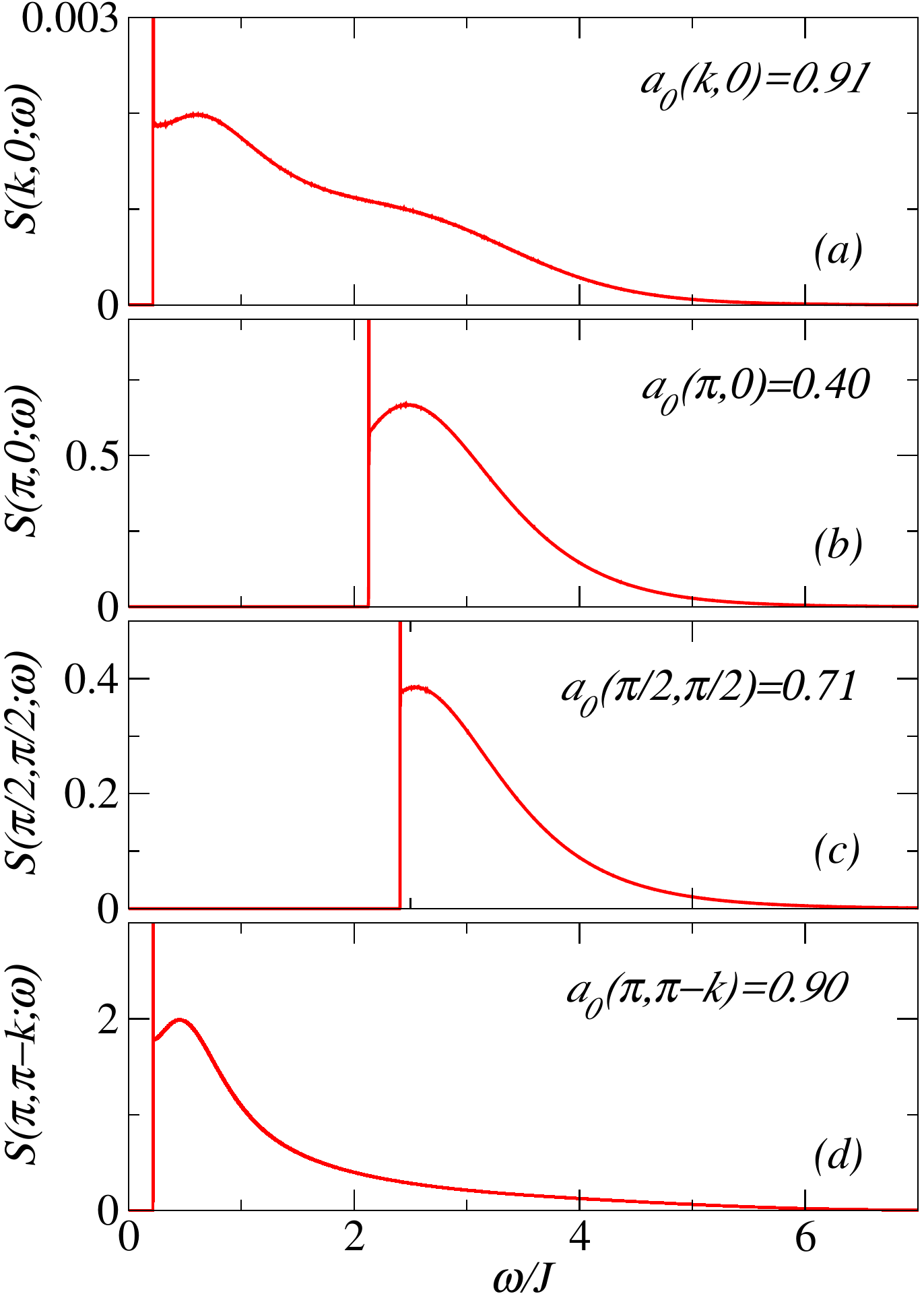}
\caption{Dynamic structure factor for $L=48$ system at four different momenta. The smallest momentum increment $2\pi/L$ is
denoted by $k$ in (a) and (d). The relative amplitude of the magnon pole is indicated in each panel.}
\label{haf-2}
\end{figure}

More detailed frequency profiles at four different wavevectors are shown in Fig.~\ref{haf-2}. In addition to the points
$(\pi,0)$ and $(\pi/2,\pi/2)$, on which many prior works have focused, results for the points closest to the gapless points
$(0,0)$ and $(\pi,\pi)$ are also shown. The results at $(\pi,0)$ and $(\pi/2,\pi/2)$ are in general in good agreement with the
previous QMC calculations \cite{sandvik01} in which the $\delta$-function contributions were also explicitly included in the
parametrization of the spectrum. The relative weight in the $\delta$-function, indicated in each panel in Fig.~\ref{haf-2}, is also in reasonably
good agreement with series expansions around the Ising limit \cite{zheng05}. The relative spectral weight of the continuum, $1-a_0({\bf q})$,
can be taken as a measure of the effect of spinwave interactions, which leads to the multi-magnon contributions often assumed to be responsible
for the continuum. We will argue later that the particularly large continuum at $(\pi,0)$ is actually due to nearly deconfined spinons.

It is not clear whether the small maximum to the right of the $\delta$-function, which we see consistently through the BZ, are real spectral
features or whether they reflect the statistical errors of the QMC data in a way similar to the most common distortion resulting from noisy
synthetic data, as seen in the tests presented in Fig.~\ref{sw}. The error level of the QMC data in all cases is a bit below $10^{-5}$, i.e.,
similar to Fig.~\ref{sw}(a). The behavior does not suggest any gap between the $\delta$-functions and the continuum.

\subsection{Finite-size effects}

It is important to investigate the size dependence of the spectral functions. For very small lattices at $T=0$, $S({\bf q},\omega)$ computed
according to Eq.~(\ref{somegasum}) for each ${\bf q}$ contains only a rather small number of $\delta$-functions and it is not possible to draw
a curve approximating a smooth continuum following a leading $\delta$-functions. Therefore, the SAC procedure does not reproduce exact Lanczos
results very well---we obtain a single broad continuum following the leading $\delta$-function, instead of several small peaks. Because the
continuum also has weight close to the leading $\delta$-function, between it and the second peak of the actual spectrum, the SAC method also
slightly underestimates the weight in the first $\delta$-function. If the continuum  emerging as the system size increases indeed is, as expected,
broad and does not exhibit any unresolvable fine-structure, the tests in Sec.~\ref{sec:sac} suggest that our methods should be able to reproduce it.

\begin{figure}[t]
\centering
\includegraphics[width=65mm]{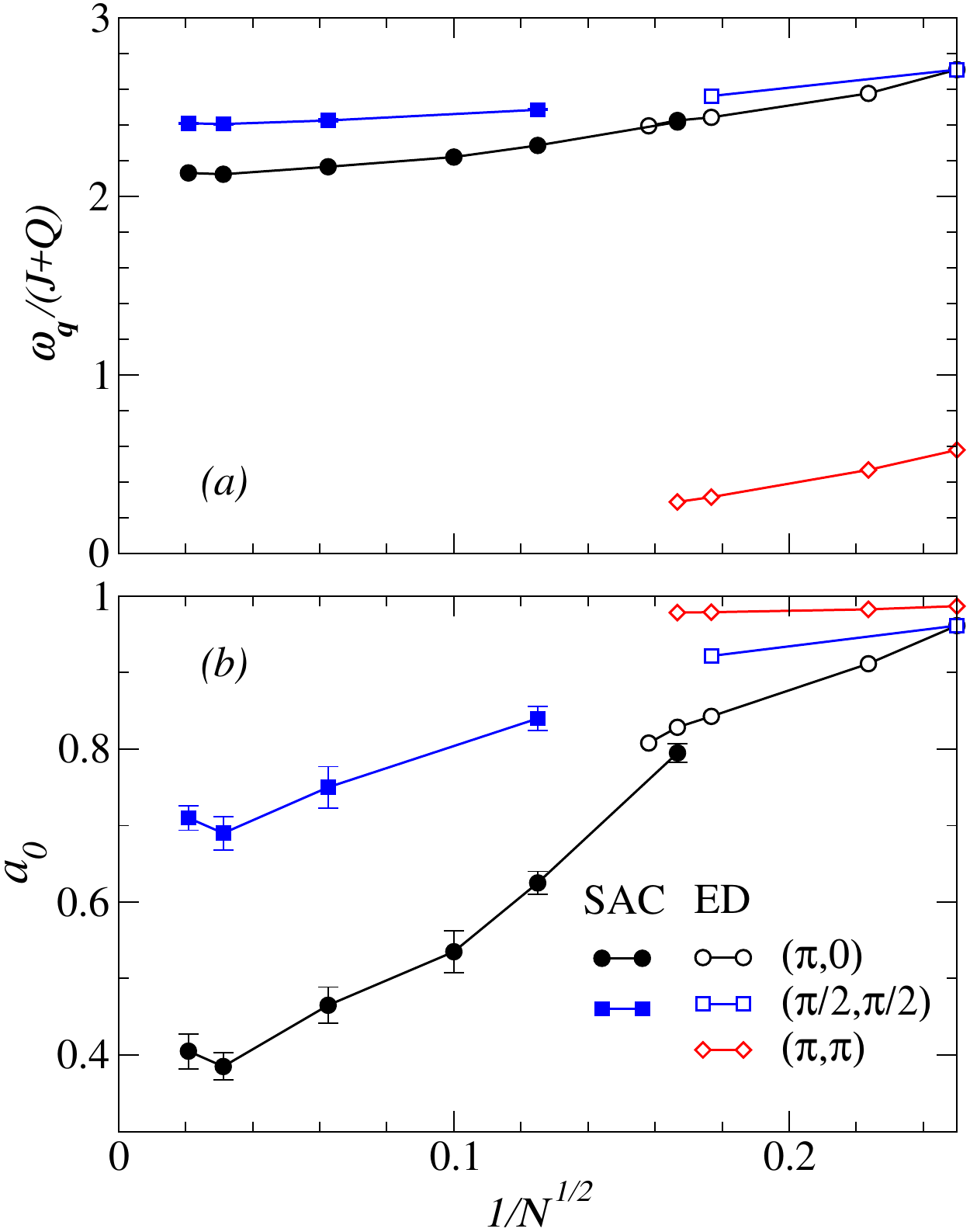}
\caption{Size dependence of the single-magnon energy (a) and weight in the magnon pole (b) at wavevectors ${\bf q}=(\pi,0)$, $(\pi/2,\pi/2)$,
  and $(\pi,\pi)$. Lanczos ED results for small systems ($L\times L$ lattices with $L=4$ and $L=6$ as well as tilted lattices
  with $N=20,32$, and $40$ sites) are shown as open circles and QMC-SAC data are presented as solid circles with error bars. The error bars were
  estimated by bootstrap analysis (i.e., carrying out the SAC procedure multiple times with random samples of the QMC data bins).}
\label{haf-6}
\end{figure}

For the $6\times 6$ lattice at ${\bf q}=(\pi,0)$, our SAC result underestimates the weight in the magnon
pole by about $5\%$, while the energy deviates by less than $1\%$. We expect these systematic errors to decrease with increasing system size,
for the reasons explained above. Fig.~\ref{haf-6} shows the size dependence of the single-magnon weight and energy at wavevectors
${\bf q}=(\pi,0)$, $(\pi/2,\pi/2)$, and $(\pi,\pi)$. At $(\pi,\pi)$ we only have Lanczos results, but even with the small
systems accessible with this method it can be seen that indeed the energy decays toward zero. The magnon weight is large, converging rapidly toward
about $97\%$, which is similar to the series-expansion result \cite{zheng05}. The energies at ${\bf q}=(\pi,0)$ and $(\pi/2,\pi/2)$ also converge rapidly,
with no detectable differences between $L=32$ and $L=48$, and a smooth transition between the ED results for small systems and QMC results for larger sizes.
The magnon weight at these wavevectors show more substantial size dependence, though again the results for the two largest sizes agree within error
bars. Here the connection between the ED and QMC results does not appear completely smooth at $(\pi,0)$, due to the difficulties for the SAC method to deal
with a spectrum with a small number of $\delta$-functions. Nevertheless, even the ED results indicate a drop in the amplitude for the larger system
sizes. The trends in $1/L$ for the QMC results suggest that the weight converges to slightly below $40\%$ at ${\bf q}=(\pi,0)$ and slightly below
$70\%$ at ${\bf q}=(\pi/2,\pi/2)$, both in very good agreement with the series-expansion results \cite{zheng05}. This agreement with a completely
different method provides strong support to the accuracy of the QMC-SAC procedures. The energies also agree very well with the previous QMC results
where particular functional forms were used to model the continuum, and the magnon amplitudes agree within $5-10\%$ (with the values indicated
in the insets of Fig.~3 in Ref.~\onlinecite{sandvik01}).

\subsection{Comparisons with experiments}

\begin{figure}[t]
\centering
\includegraphics[width=75mm]{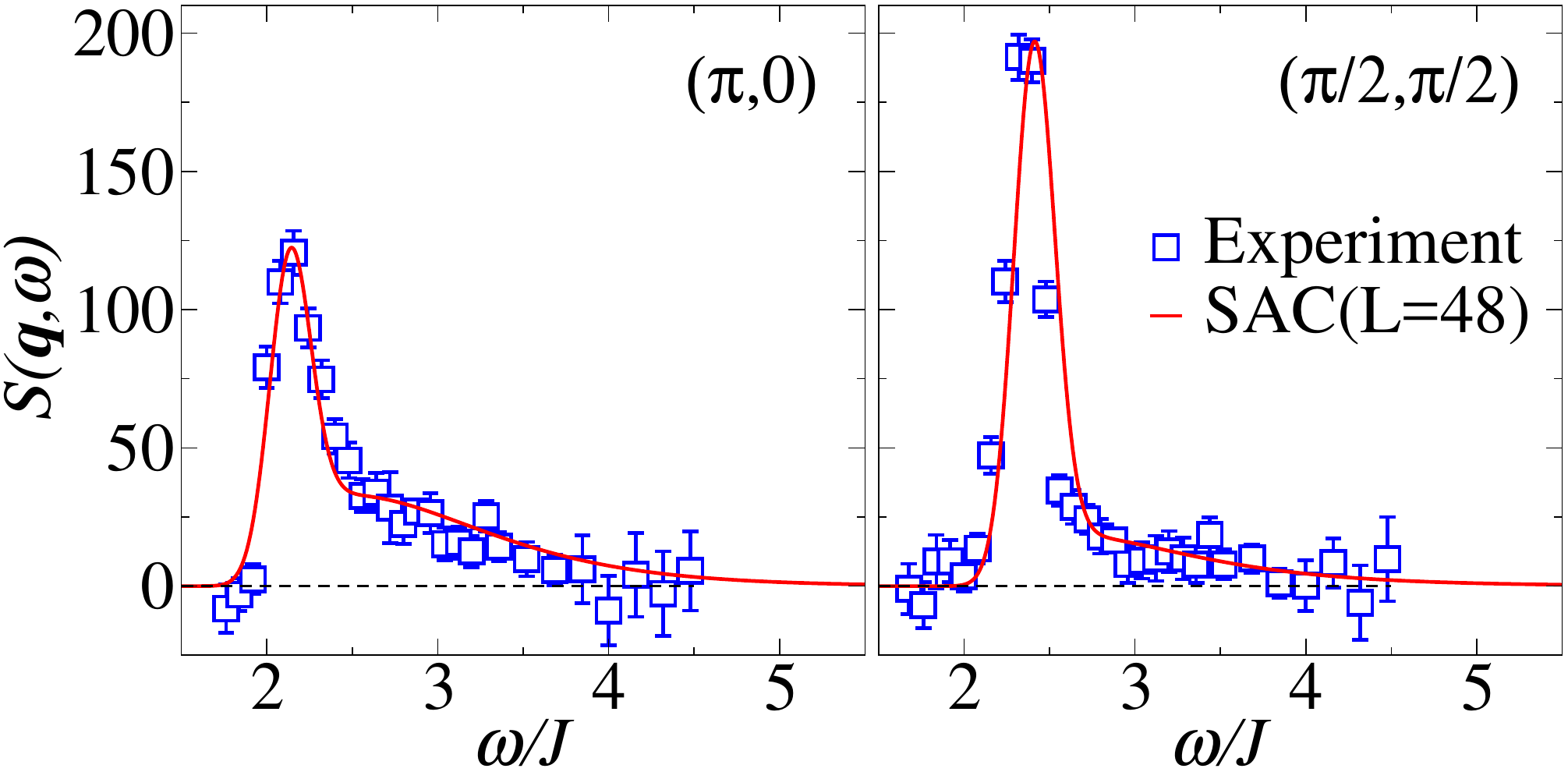}
\caption{Comparison of the CFTD experimental data \cite{piazza15} (the full scattering cross section corresponding to unpolarized neutrons)
 and our QMC-SAC spectral functions at wavevectors ${\bf q}=(\pi,0)$ and ${\bf q}=(\pi/2,\pi/2)$.
  To account for experimental resolution, we have convoluted the QMC-SAC spectral functions in Figs.~\ref{haf-2}(b,c) with a common Gaussian broadening
  (half-width $\sigma=0.12J$). We have renormalized the exchange constant by a factor $1.02$ relative to the original value in Ref.~\onlinecite{piazza15}, and
  to match the arbitrary factor in the experimental data we have further multiplied both of our spectra by a factor $\approx 50$.}
\label{haf-3}
\end{figure}

In the discussion of the recent neutron-scattering experiments on CFTD \cite{piazza15}, it was argued that the large continuum in the
$(\pi,0)$ spectrum is due to fully deconfined spinons, and a variational RVB wavefunction was used to support this interpretation. We will
discuss our different picture of nearly deconfined spinons further in Sec.~\ref{sec:heff}. Here we first compare the $(\pi,0)$ and $(\pi/2,\pi/2)$
results with the experimental data without invoking any interpretation. The experimental scattering cross section in Ref.~\onlinecite{piazza15}
was shown versus the frequency $\omega/J$ normalized by the estimated value of the coupling constant ($J \approx 6.11$ meV). Keeping the same scale,
we should only convolute our spectral functions with an experimental Gaussian broadening. We optimize this broadening to match the data and find
that a half-width $\sigma=0.12J$ of the Gaussian works well for both wavevectors---which is the same as the instrumental broadening reported
for the experiment \cite{piazza15}. Since the neutron data are presented with an arbitrary scale for the scattering intensity we also have to multiply
our $S({\bf q},\omega)$ for each ${\bf q}$ by a common factor. The agreement with the data at both $(\pi,0)$ and $(\pi/2,\pi/2)$ is very good, and
can be further improved by dividing $\omega/J$ in the experimental data by $1.02$, which corresponds to $J \approx 6.23$ meV, which should still
be within the errors of the experimentally estimated value. As shown in Fig.~\ref{haf-3}, the agreement with the experiments is not perfect but
probably as good as could possibly be expected, considering small effects of the weakly ${\bf q}$-dependent form factor \cite{formfactor}
and some influence of weak interactions beyond $J$ (longer-range exchange, ring exchange, spin-phonon couplings, disorder, etc.).

\begin{figure}[t]
\centering
\includegraphics[width=70mm]{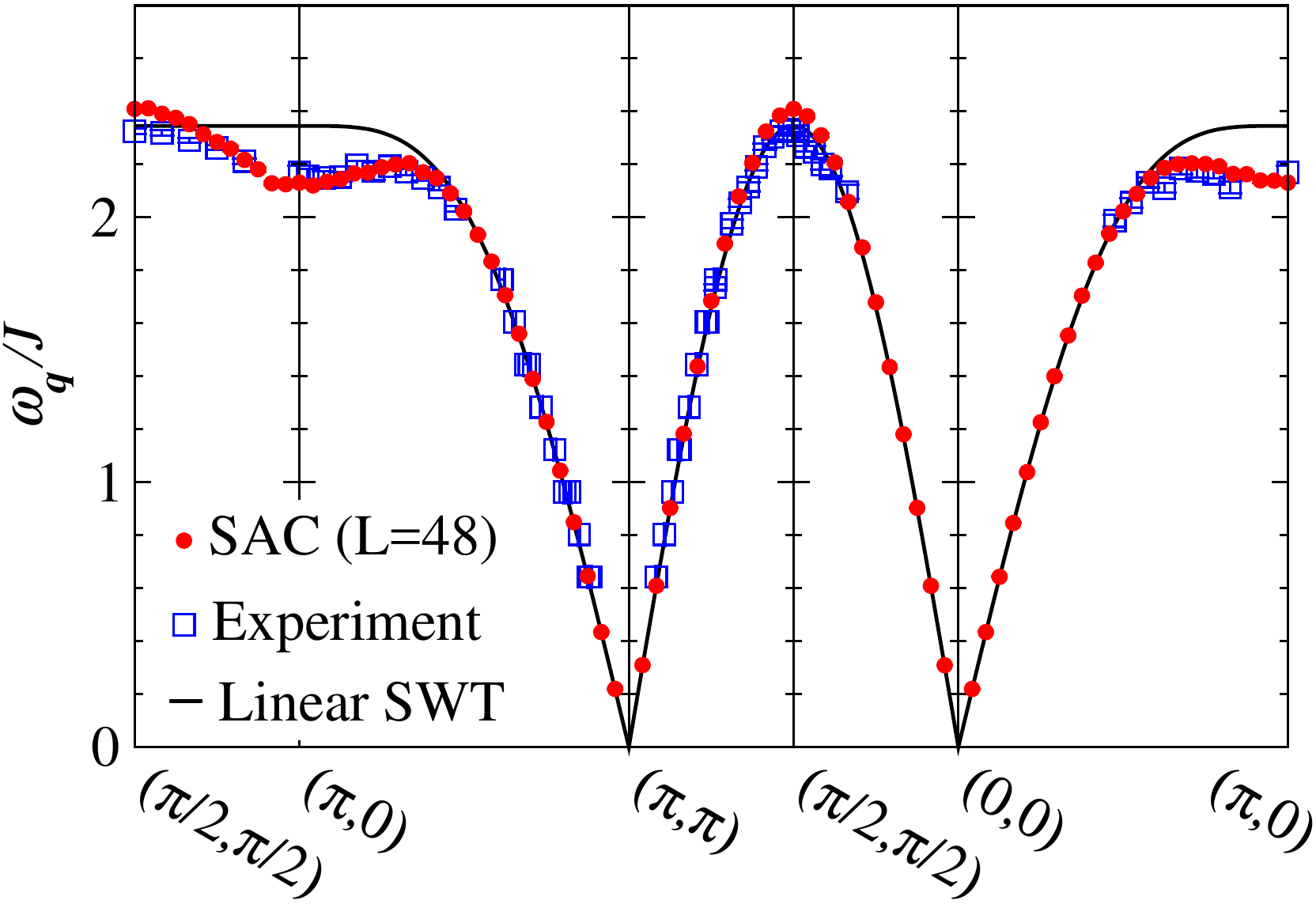}
\caption{Single-magnon dispersion $\omega_{\bf q}$ along a representative path of the magnetic BZ. The CFTD experimental data from Ref.~\onlinecite{piazza15}
are shown as blue squares and the QMC-SAC data (the location of the magnon pole) are shown with red circles. We also show the linear SWT dispersion
(black curve) adjusted by a common factor corresponding to the exact spinwave velocity  $c=1.65847$ \cite{sen15}.} 
\label{haf-4}
\end{figure}

The single-magnon dispersion, the energy $\omega_{\bf q}$ in Eq.~(\ref{Sfunction}), is compared with the corresponding experimental peak
position in Fig.~\ref{haf-4}. The linear spinwave dispersion is shown as a reference, using the best available value of the renormalized
velocity $c=1.65847$ \cite{sen15}. Our results agree very well with the spinwave dispersion at low energies, and with the experimental CDFT
data \cite{piazza15} also in the high-energy regions where the spinwave results are not applicable. The only statistically significant
deviation, though rather small, is at ${\bf q} \approx (\pi/2,\pi/2)$, where the experimental energy is lower (as seen also in the peak
location in Fig.~\ref{haf-3}). Still, overall, one must conclude that CFTD is an excellent realization of the square-lattice Heisenberg model
at the level of current state-of-the-art experiments. It would certainly be interesting to improve the frequency resolution further and try to
analyze higher-order effects, which should become possible in future neutron scattering experiments.

\subsection{Wavevector dependence of the single-magnon amplitude}

\begin{figure}[t]
\centering
\includegraphics[width=70mm]{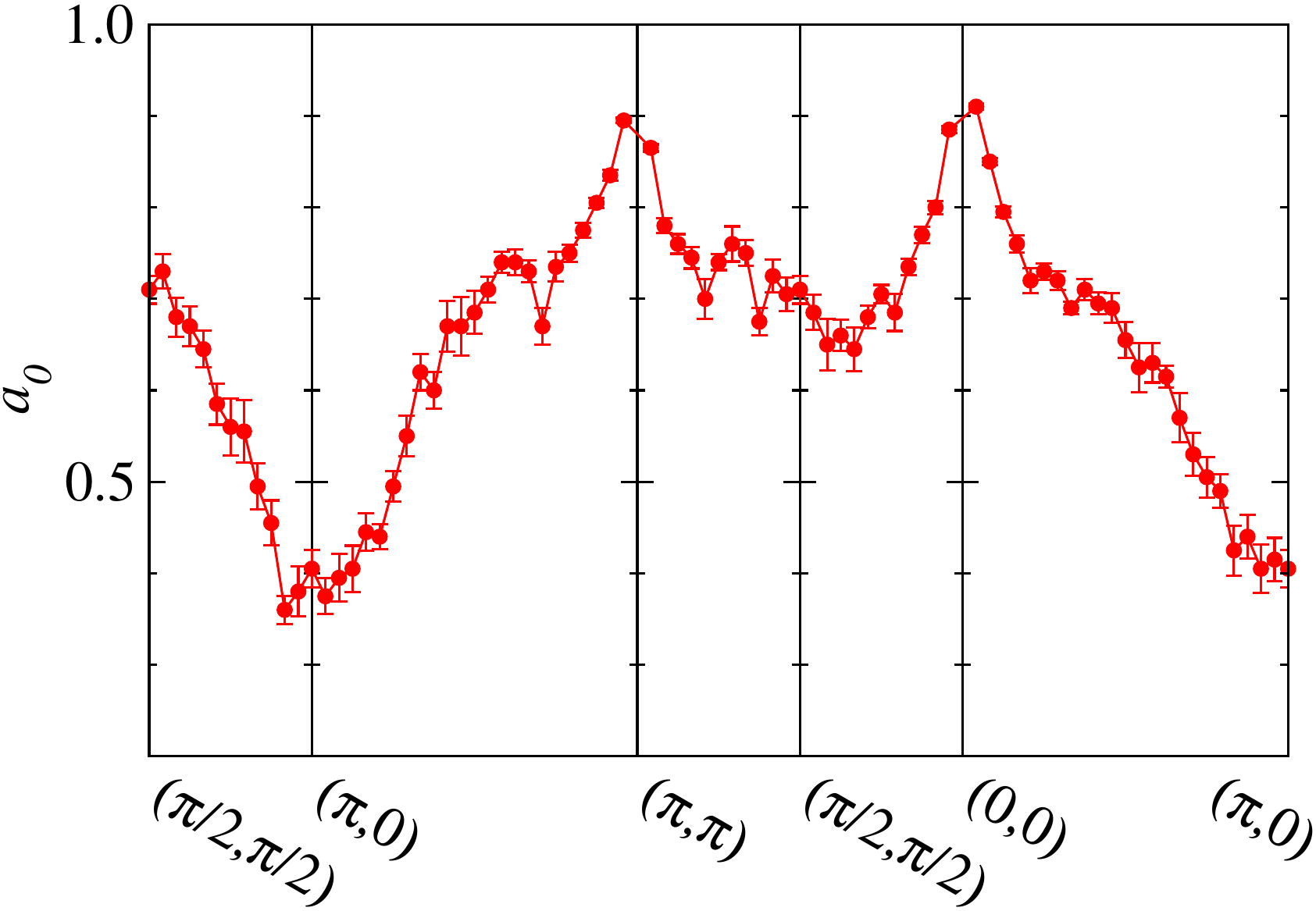}
\caption{Relative spectral weight of the single-magnon pole along the representative path in the BZ for the $L=48$ Heisenberg system.
 Error bars were estimated by bootstrapping.}
\label{haf-5}
\end{figure}

We next look at the variation of the relative magnon weight $a_0({\bf q})$ along the representative path of the BZ for $L=48$,
shown in Fig.~\ref{haf-5}. For ${\bf q}\to(0,0)$ and $(\pi,\pi)$ the weight $a_0$ increases and appears to tend close
to $1$. From the results exactly at $(\pi,\pi)$ in Fig.~\ref{haf-6} we know that in this case the remaining weight in the continuum
should be about $3\%$, which is also in good agreement with the series results in Ref.~\onlinecite{zheng05}, where a similar non-zero multi-magnon
weight was also found as $q \to 0$. At ${\bf q}=(\pi/2,\pi/2)$, as also shown in Fig.~\ref{haf-6}, the magnon pole contains about
$70\%$ of the weight, while at ${\bf q}=(\pi,0)$ this weight is reduced to about $40\%$. Both of these are also in good agreement with
Ref.~\onlinecite{zheng05}, and in fact throughout the BZ path we find no significant deviations from the series results. This again reaffirms
the ability of the SAC procedure to correctly optimize the amplitude of the leading $\delta$-function. It should be noted that the series
expansion around the Ising model does not produce the full spectral functions, only the single-magnon dispersion and weight.

The depletion seen in Fig.~\ref{haf-5} of the single-magnon weight in a neighborhood of ${\bf q}=(\pi,0)$ can also be
related to the experimental data for CFTD. In Fig.~1(a) of Ref.~\onlinecite{piazza15}, a color coding is used for the scattering
intensity such that even a modest reduction in the coherent single-magnon weight has a large visual impact. The region
in which the spectral function is smeared out with no sharp feature in this representation corresponds closely to the
region where the single-magnon weight drops from about $60\%$ to $40\%$ in our Fig.~\ref{haf-5}.

\subsection{Alternative ways of analytic continuation}

One could of course argue that the existence of the magnon pole at $(\pi,0)$ is not proven by our calculations since it has
been built into our parametrization of the spectral function. While it is clear that our approach cannot distinguish between a very
narrow peak and a $\delta$-function, if the broadening is significant for some ${\bf q}$, so that the main peak essentially becomes
part of the continuum, we would expect the optimal amplitude $a_0({\bf q})$ to be very small or vanish. Nevertheless, to explore the
possibility of spectra without magnon pole, we also have carried out the analytic continuation in two alternative ways, using
the parametrization in Fig.~\ref{deltas}(a) without special treatment of the lowest frequency, or by imposing a lower frequency
bound.

\begin{figure}[t]
\centering
\includegraphics[width=70mm]{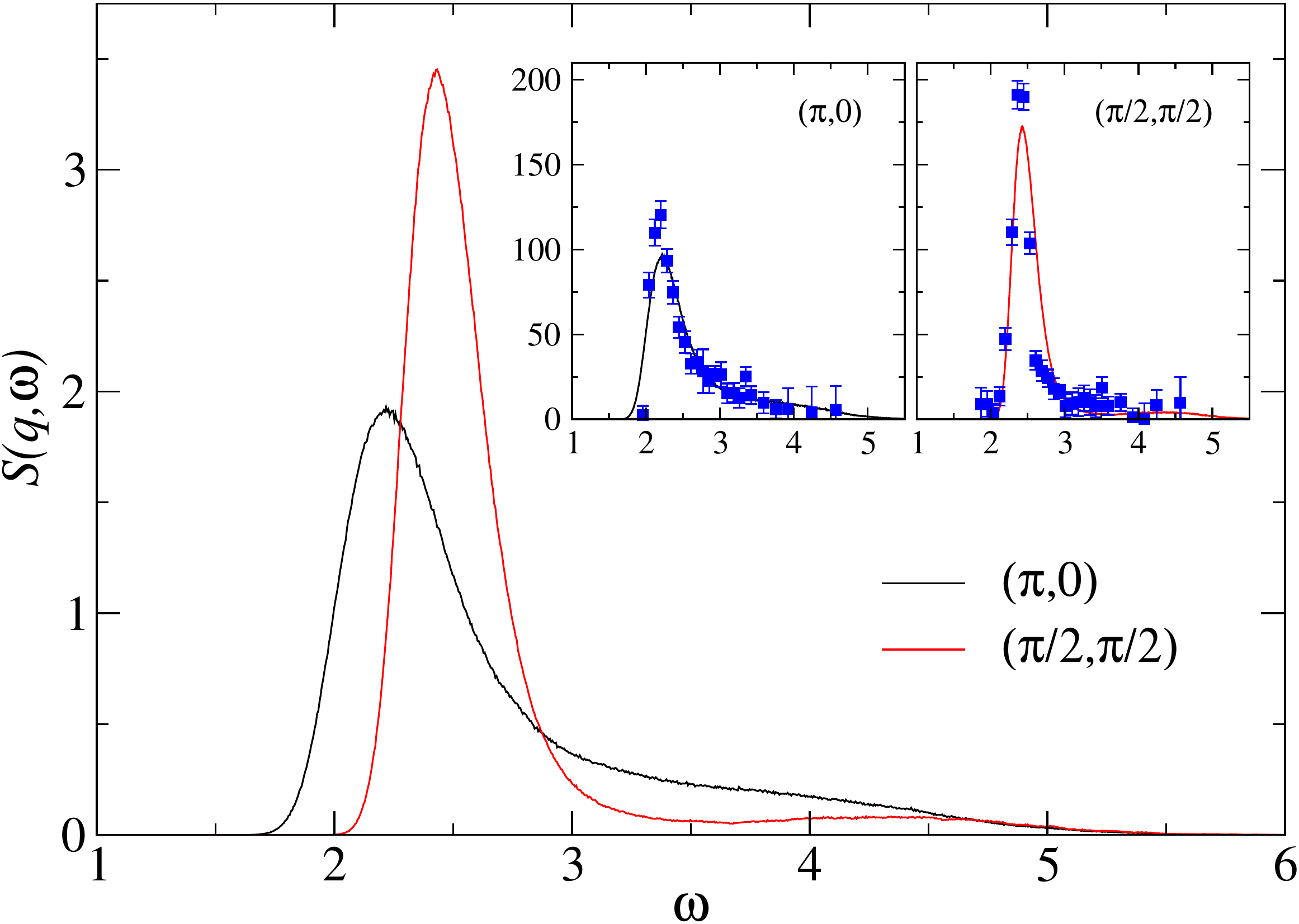}
\caption{Spectral functions at ${\bf q}=(\pi,0)$ and $(\pi/2,\pi/2)$ obtained using unconstrained SAC with the
  parametrization in Fig.~\ref{deltas}(a). The insets show comparisons with the experimental data \cite{piazza15}, where
  we have only adjusted a common amplitude to match the areas under the peaks.}
\label{sqw0}
\end{figure}

Sampling without any constraints with $N_\omega=1000$
$\delta$-functions gives the results at $q=(\pi,0)$ and $(\pi/2,\pi/2)$ shown in Fig.~\ref{sqw0}. Here one can distinguish a peak
in each case in the general neighborhood of where the $\delta$-function is located in Figs.~\ref{haf-2}(b,c), with the the maximum shifted
slightly to higher frequencies and weight extending significantly to lower frequencies. At
${\bf q}=(\pi/2,\pi/2)$ there is now a shallow minimum before a low broad distribution at higher energies. This kind of behavior is typical
for analytic continuation methods when there is too much broadening at low frequency, which leads to a compensating (in order to match
the QMC data) depletion of weight above the main peak. Similarly, the up-shift of the location of the peak frequency at both {\bf q}
relative to Fig.~\ref{haf-2} is due to there being weight also at $\omega < \omega_{\bf q}$ where there should be none or much less weight.
In the insets of Fig.~\ref{sqw0} we show comparisons with the CFTD experimental data. Here the SAC spectral functions are broader than
the experimental profiles and we have not applied any additional broadening. It is clear that the SAC results here do not match the
experiments as well as in Fig.~\ref{haf-3}, most likely because the QMC data are no sufficiently precise to reproduce a narrow magnon
pole, thus also leading to other distortions at higher energy.

\begin{figure}[t]
\centering
\includegraphics[width=70mm]{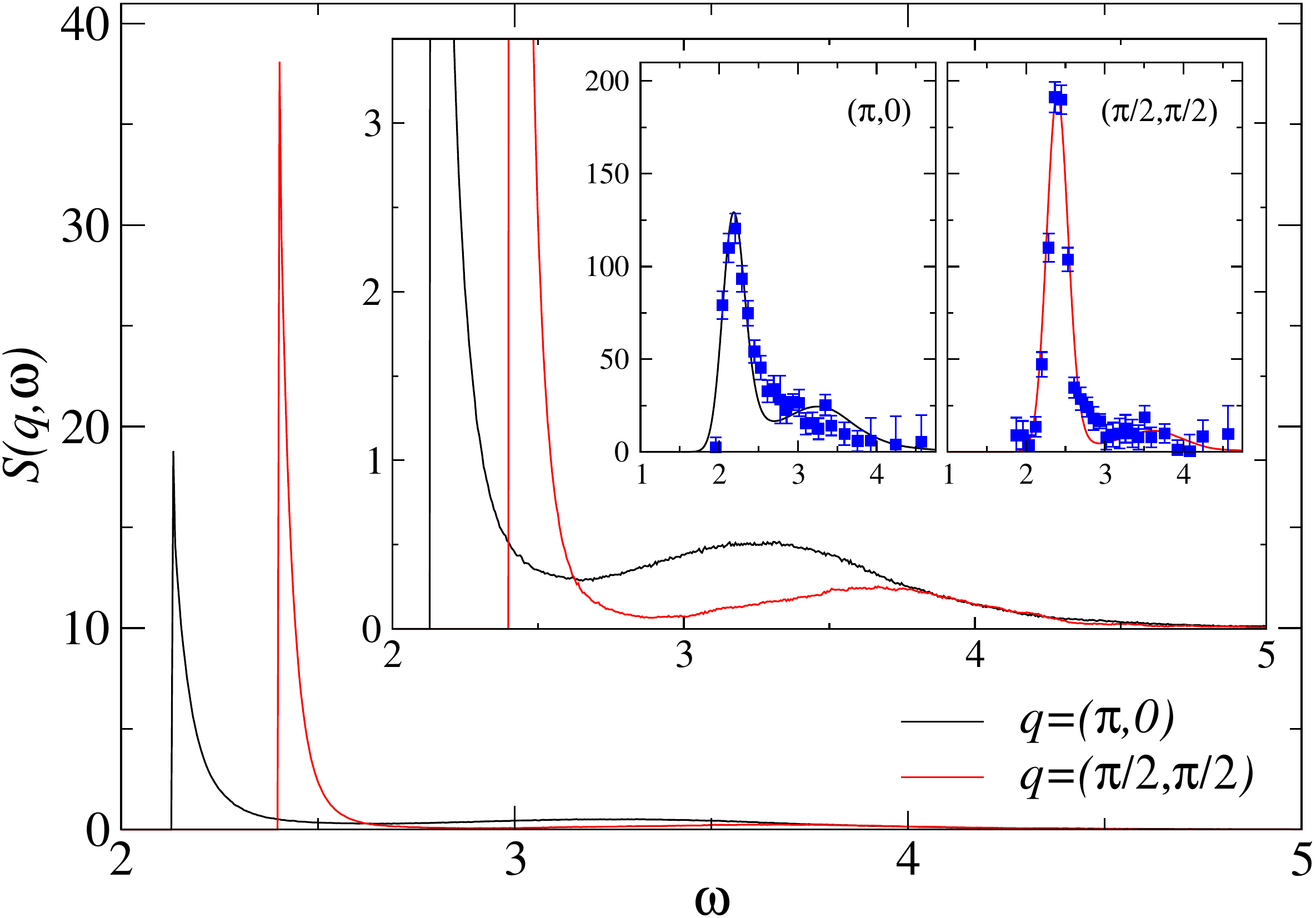}
\caption{Spectral functions obtained using sampling with the parametrization in
  Fig.~\ref{deltas}(a) under the constraint that no weight falls below the lower bounds determined with a $\delta$-function
  at the lower edge (Fig.~\ref{haf-2}); $\omega_q=2.13$ and $2.40$ for ${\bf q}=(\pi,0)$ and $(\pi/2,\pi/2)$, respectively.
  The inset shows the results on a different scale to make the continua better visible. The insets of the inset show comparisons
  with the experimental data, where we have broadened the numerical results by Gaussian convolution and adjusted a common amplitude.}
\label{sqw2}
\end{figure}

In order to reduce the broadening and other distortions arising as a consequence of spectral weight spreading out in the SAC sampling
procedure due to entropic pressure \cite{sandvik16} into regions where there should be no weight, we also carried out SAC runs with the
constraint that no $\delta$-function can go below the lowest energy determined with the dominant $\delta$-function present. These energies,
$\omega_q = 2.13$ and $2.40$ for ${\bf q}=(\pi,0)$ and $(\pi/2,\pi/2)$, respectively, are in excellent agreement with the series expansions around
the Ising limit \cite{zheng05} and, in the case of $(\pi/2,\pi/2)$, also with the well-converged high-order spin-wave expansion
\cite{Igarashi92,Canali93,Igarashi05,syromyatnikov10}. There is therefore good reason to trust these as being close to the actual
energies. As seen in Fig.~\ref{sqw2}, there is a dramatic effect of imposing the lower bound---the main peak is much higher and narrower
than in Fig.~\ref{sqw0} and an edge is formed at $\omega_q$. Most likely the peaks are still broadened on the right side, 
and again this broadening has as a consequence a local minimum in spectral weight before a broad second peak, which is now seen 
for both ${\bf q}$ points. In this case the comparisons with the experiments (insets of Fig.~\ref{sqw2}) is overall somewhat better than
with the completely unconstrained sampling in Fig.~\ref{sqw0}, but still we see signs of a depletion of spectral weight to the right of
the main peak that is not present in the experimental data. We take the $\omega_0$-constrained spectra as upper limits in terms of the
widths of the main magnon peaks, and most likely the true spectra are much closer to those obtained with the optimized $\delta$-functions
in Fig.~\ref{haf-2}.

In summary, the results of these alternative ways of carrying out the SAC process reaffirm that there indeed should be a leading very narrow
magnon pole, close to a $\delta$-function, at both ${\bf q}=(\pi,0)$ and $(\pi/2,\pi/2)$. While the pole strictly speaking may have some damping,
our good fits with a pure $\delta$-function in Fig.~\ref{haf-3} indicates that such damping should be extremely weak, as also expected on
theoretical grounds \cite{chernyshev06,chernyshev09}.

\section{$\text {J-Q}$ model}
\label{sec:jq}

The AFM order parameter in the ground state of the Heisenberg model is significantly reduced by zero-point quantum fluctuations from its classical
value $m_s=1/2$ to about $0.307$ \cite{Reger88,sandvik10a}. It can be further reduced when frustrated interactions are included, eventually leading
to a quantum-phase transition into a non-magnetic state, e.g., in the frustrated $J_1$-$J_2$ Heisenberg model 
\cite{dagotto89,gelfand89,hu13,gong14,morita15,wang17}. In the
$J-Q$ model \cite{sandvik07}, the quantum phase transition driven by the four-spin coupling $Q$ appears to be a realization of the deconfined
quantum critical point \cite{shao16}, which separates the AFM state and a spontaneously dimerized ground state; a columnar VBS. The model is amenable
to large-scale QMC simulations and we consider it here in order to investigate the evolution of the dynamic structure factor upon reduction of
the AFM order and approaching spinon deconfinement.

The $J$-$Q$ Hamiltonian can be written as
\cite{sandvik07}, 
\begin{equation}
H = -J \sum_{\langle ij\rangle} P_{ij} -Q \sum_{\langle ijkl\rangle} P_{ij}P_{kl},
\label{jqham}
\end{equation}
where $P_{ij}$ is a singlet projector on sites $ij$,
\begin{equation}
P_{ij} = 1/4 - {\bf S}_i \cdot {\bf S}_j,
\end{equation}
here on the nearest-neighbor sites. In the four-spin interaction $Q$ the site pairs $ij$ and $kl$ form horizontal and vertical edges
of $2\times 2$ plaquettes. All translations and $90^\circ$ rotation of the operators are included in Eq.~(\ref{jqham}) so that all
the symmetries of the square lattice are preserved.

\begin{figure}[t]
\centering
\includegraphics[width=65mm]{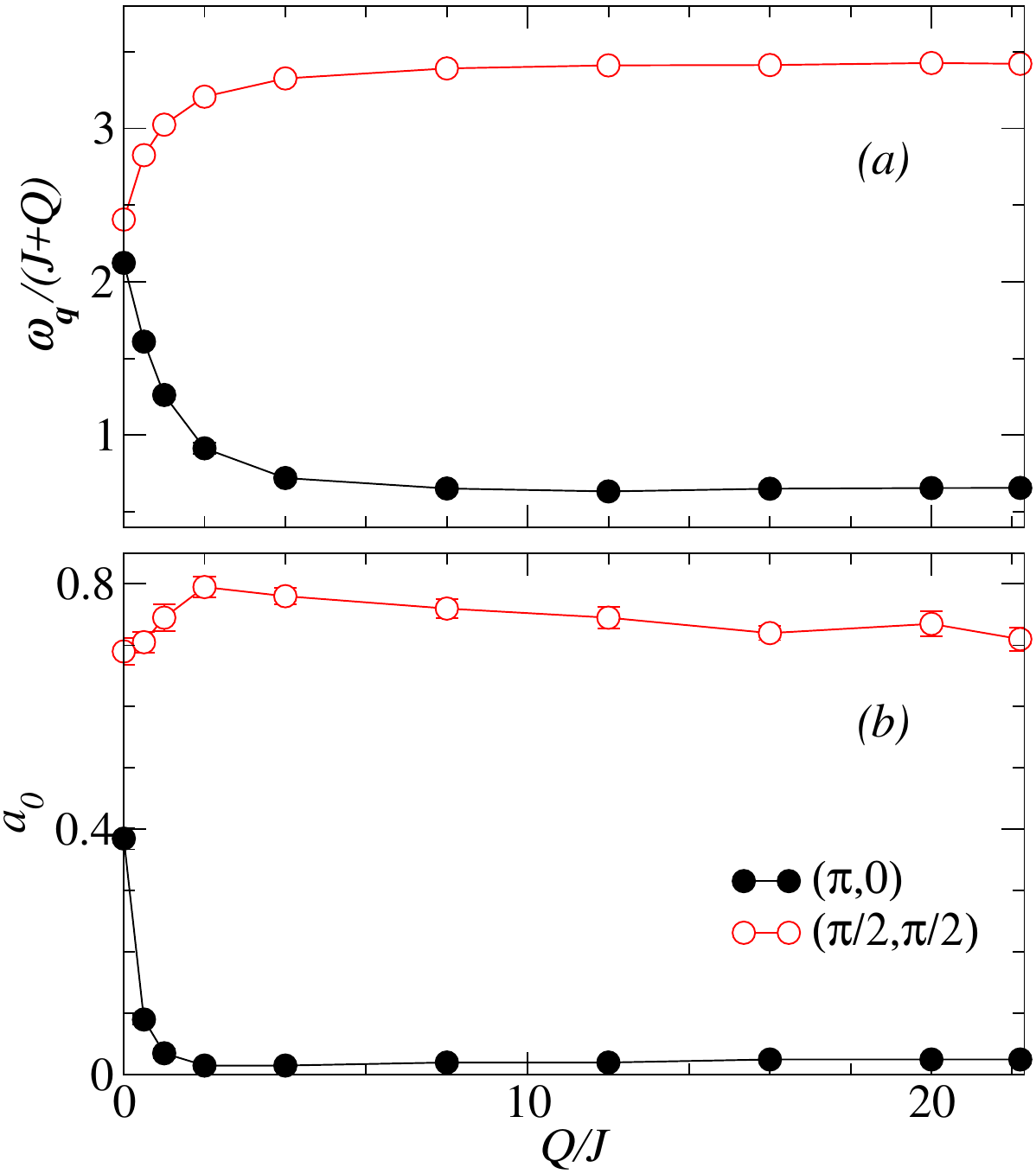}
\caption{Results for the $J$-$Q$ model at ${\bf q}=(\pi,0)$ and $(\pi/2,\pi/2)$, calculated on the $L=32$ lattice. 
  The lowest excitation energy $\omega_q$ (a) and the relative weight of the single-magnon contribution (b) are shown as functions
  of the coupling ratio $Q/J$ from the Heisenberg limit ($Q/J=0$) to the deconfined quantum critical point ($Q_c/J\approx 22$).} 
\label{jq}
\end{figure}

In addition to strong numerical evidence of a continuous AFM--VBS transition in the $J$-$Q$ model (most recently in Ref.~\onlinecite{shao16}),
there are also results pointing directly to spinon excitations at the critical point, in accord with the scenario of deconfined quantum criticality
\cite{senthil04a,senthil04b} (where, strictly speaking, there may be weak residual spinon-spinon interactions, though those may only be important
in practice only at very low energies \cite{tang13}). Moreover, the set of gapless points is expanded from just the points ${\bf q}=(0,0)$ and
$(\pi,\pi)$ in the N\'eel state to also ${\bf q}=(\pi,0)$ and $(0,\pi)$ \cite{spanu06,suwa16} at the critical point. Recent results point to
linearly dispersing spinons with a common velocity around all the gapless points \cite{suwa16}.

Here our primary aim is to study how the magnon poles and continua in $S({\bf q},\omega)$ at ${\bf q}=(\pi,0)$ and $(\pi/2,\pi/2)$ evolve as the coupling
ratio $Q/J$ is increased. We use the same SAC parametrization as in the previous section, with a leading $\delta$-function whose amplitude is optimized by
finding the minimum in $\langle \chi^2\rangle$ versus $a_0({\bf q})$. We first consider the $L=32$ lattice and show our results for the energy and the
relative amplitude in Fig.~\ref{jq} as functions of the coupling ratio $Q/J$ all the way from the Heisenberg limit to the deconfined quantum critical
point. Here the most notable aspect is the rapid drop in the magnon weight at ${\bf q}=(\pi,0)$, even for small values of
$Q/J$, while at ${\bf q}=(\pi/2,\pi/2)$ the weight stays large,  $70-80\%$, over the entire range. The energies depend on the normalization
and here we have chosen $J + Q$ as the unit. We know from past work that the ${\bf q}=(\pi,0)$ energy at $Q_c/J$ vanishes in the thermodynamic limit
but the reduction in the finite-size gap with the system size is rather slow \cite{suwa16}, and for the $L=32$ lattice considered here we are still
far from the gapless behavior.

\begin{figure}[t]
\centering
\includegraphics[width=65mm]{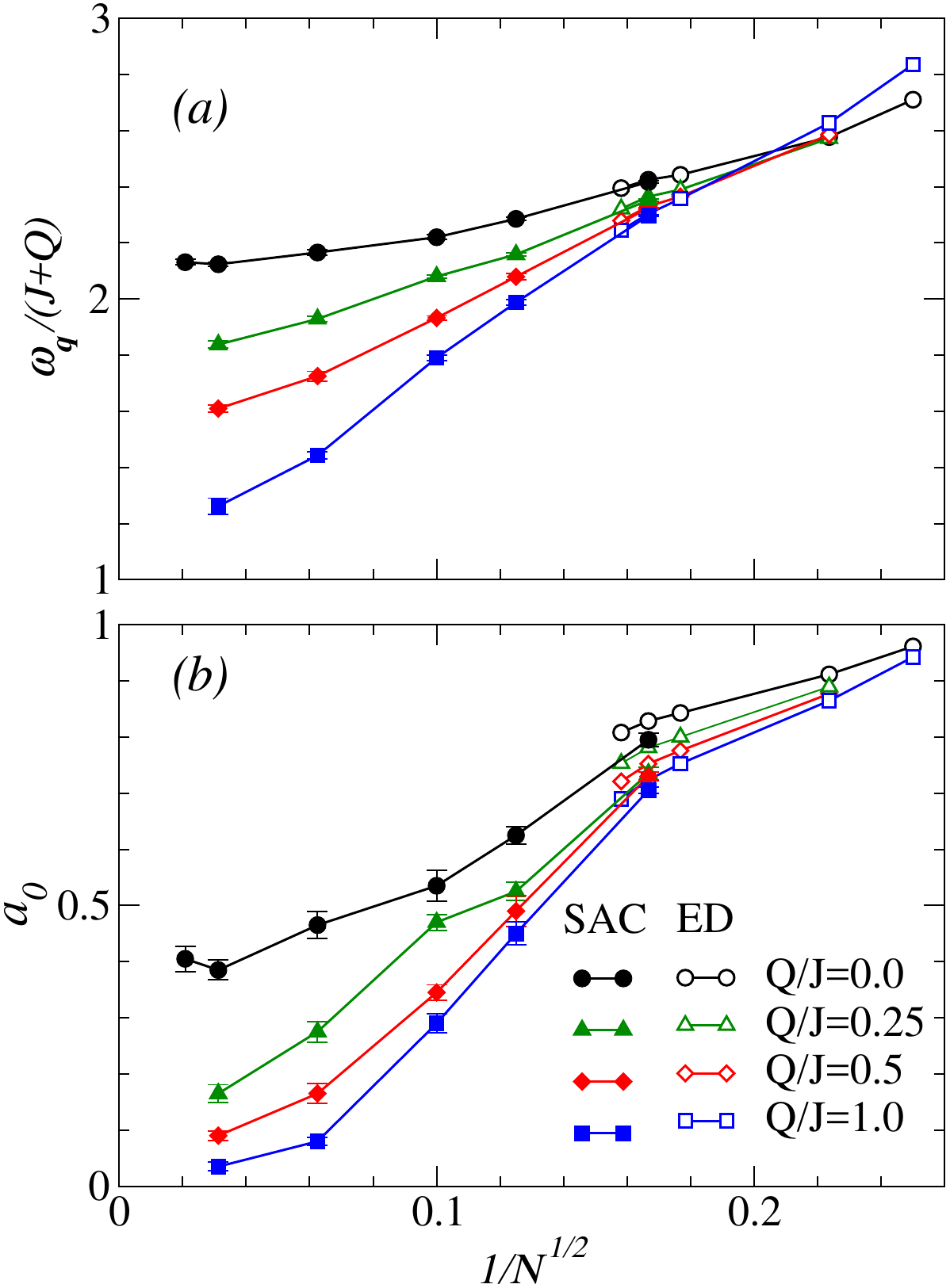}
\caption{Size dependence of the excitation energy $\omega_{\bf q}$ (a) and the relative weight of the magnon pole $a_0({\bf q})$ (b)
  at ${\bf q}=(\pi,0)$ close to the Heisenberg limit of the $J$-$Q$ model.} 
\label{fss}
\end{figure}

We focus on the effects on small $Q$, where reliable extrapolations to infinite size are possible, and show the size dependence of the lowest excitation energy 
and the magnon amplitude at ${\bf q}=(\pi,0)$ for several cases in Fig.~\ref{fss}. We again show Lanczos ED results for small systems and QMC-SAC
results for larger sizes. For the only common system size, $L=6$, the energies agree very well, as in the pure Heisenberg case discussed in the previous
section, while the QMC-SAC calculations underestimate the magnon weight by a few percent due to the inability to resolve the details of a spectrum
consisting of just a small number of $\delta$-functions. The most interesting feature is the dramatic reduction in the magnon weight even for very small
ratios $Q/J$. For $Q/J=0.25$ and $0.5$, the size dependence indicates small remaining magnon poles, while at $Q/J=1$ it appears that the $\delta$-function
completely vanishes in the thermodynamic limit.

\begin{figure}[t]
\centering
\includegraphics[width=70mm]{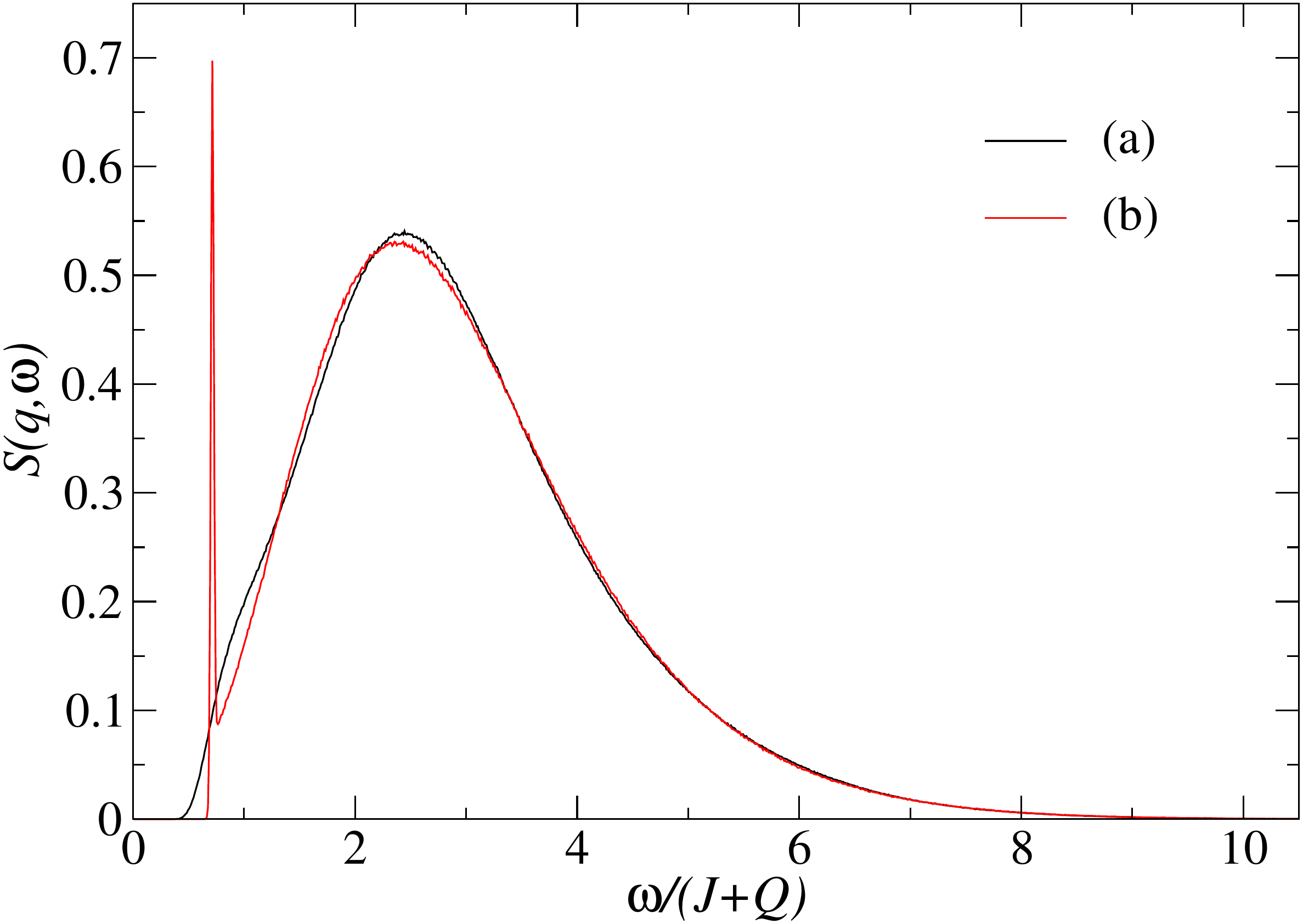}
\caption{The ${\bf q}=(\pi,0)$ dynamic structure factor of the $J$-$Q$ model at $Q/J=4$ obtained using SAC with the two parametrizations
  of the the spectrum in Figs.~\ref{deltas}(a,b). The relative weight of the leading $\delta$-function in (b) is $1.4\%$.}
\label{sqw4}
\end{figure}

In Fig.~\ref{sqw4} we show the full ${\bf q}=(\pi,0)$ dynamic structure factor at $Q/J=4$, obtained with  both the parametrizations in Fig.~\ref{deltas}.
The optimal weight of the leading $\delta$-function is only $1.4\%$ for this $L=32$ lattice, and the finite-size behavior indicates that no magnon pole at all
should be present in the thermodynamic limit in this case. When no leading $\delta$-function is included in the SAC treatment, i.e., with unrestricted
SAC sampling with the parametrization in Fig.~\ref{deltas}(a), there is a little shoulder close to where the $\delta$-function is located with the other
parametrization. The differences at higher frequencies are very minor. This is very different from the large change in the entire spectrum when
unrestricted sampling is used for the same wavevector in the pure Heisenberg model, Fig.~\ref{sqw0}, which is clearly because of the much larger magnon
pole in the latter case. This comparison also reinforces the ability of our SAC method to extract the correct weight of the leading $\delta$-function.

These results for the $J$-$Q$ model show that the magnon picture at $q=(\pi,0)$ fails even with a rather weak deformation of 
the Heisenberg model. Thus, it seems likely that the reduced excitation energy and coherent single-magnon weight at $q=(\pi,0)$,
observed in the Heisenberg model as well as experimentally in CFTD, is a precursor to deconfined quantum criticality. If that is indeed
the case, then it may be possible not only to describe the continuum in $S({\bf q},\omega)$ around ${\bf q}=(\pi,0)$ in terms of spinons
\cite{piazza15}, but also to characterize the influence of spinons on the remaining sharp magnon pole. We next consider a simple effective
Hamiltonian to address this possibility.

\section{Nature of the excitations}
\label{sec:heff}

Motivated by the numerical results presented in Secs.~\ref{sec:hberg} and \ref{sec:jq}, we here propose a mechanism 
of the excitations in the square-lattice Heisenberg model where the magnons have an internal structure corresponding to
a mixing with spinons at higher energy. Our physical picture is that the magnon resonates in and out of the spinon space,
which, in the absence of spinon-magnon couplings, exists above the bare magnon energy. We will construct a simple effective
coupled magnon-spinon model describing such a mechanism. The model resembles the simplest model for the exciton-polariton
problem, where the mixing is between light and a bound electron-hole pair (exciton). Here a bare photon can be absorbed by
generating an exciton, and subsequently the electron and hole can recombine and emit a photon. This resulting collective
resonating electron-hole-photon state is called an exciton-polariton \cite{hopfield58,mahan}. The spinon-magnon model introduced
here is more complex, because the magnon interacts not just with a single bound state but with a whole continuum of spinon states
with or without (depending on model parameters) spinon-spinon interactions.

We start below by discussing the dispersion relations of the bare magnon and spinons, and then present details of the mixing process
and the effective Hamiltonian. We will show that the model can reproduce the salient spectral features found for the Heisenberg
and $J$-$Q$ models in the preceding section, in particular the differences between wavevectors $(\pi,0)$ and $(\pi/2,\pi/2)$ and
the evolution of the spectral features when the $Q$ interaction is turned on, which in the effective model corresponds to
lowering the bare spinon energy.

\subsection{Effective Hamiltonian}

In spinwave theory, the excitations of the square-lattice Heisenberg antiferromagnet are described as magnons, which to order
$1/S$ disperse according to
\begin{equation}
\omega^{\text m}({\bf q})=c^{\text m}\sqrt{2-\frac{1}{2}\left[\cos(q_x)+\cos(q_y)\right]^2},
\label{m-dispersion}
\end{equation}
where $c^{\text m}$ is the spin wave velocity (the value of which is $c^{\text m}=1.637412$ when calculated to this order). We will take
this form of $\omega^{\text m}({\bf q})$ as the bare magnon energy in our model but treat the velocity as an adjustable bare parameter.

Spinons are well understood in the $S=1/2$ AFM Heisenberg chain, where the dispersion relation is \cite{faddeev81,moller81}
\begin{equation}
\omega(k)=\frac{\pi}{2}\sin(k),
\label{1d-omega}
\end{equation}
and an $S=1$ excitation with wavenumber ${q}$ can exist at all energies $\omega({k}_1)+\omega({k}_2)$ with ${k}_1+{k}_2={q}$.
In 2D, we use as input results of a recent QMC study of the excitation spectrum at the deconfined quantum critical point of the J-Q model \cite{suwa16},
where four gapless points at ${\bf q}=(0,0),(\pi,0),(0,\pi)$, and $(\pi,\pi)$ were found in the $S=1$ excitation spectrum (confirming a general expectation
of a system at a continuous AFM--VBS transition \cite{spanu06}). This dispersion relation is interpreted as the lower bound of a two-spinon continuum,
which should also be the dispersion relation for a single spinon. In the effective model we will use the simplest spinon dispersion relation with
the above four gapless points and shape in general agreement with the findings in Ref.~\onlinecite{suwa16},
\begin{equation}
\omega^\text{s}({\bf q})=c^{\text s}\sqrt{1-\cos^2(q_x)\cos^2(q_y)},
\label{free-s-dispersion}
\end{equation}
which can also be regarded as a 2D generalization of the 1D spinon dispersion, Eq.~(\ref{1d-omega}). The common velocity $c^{\text s}$ at the gapless
points was determined for the critical $J$-$Q$ model \cite{suwa16} but here we will regard it as a free parameter.

One of our basic assumptions will be that spinons exist in the system also in the AFM phase, but they are no longer gapless and interact with
the magnon excitations. We will add a constant $\Delta$ to the spinon energy Eq.~(\ref{free-s-dispersion}) to model the evolution of the bare spinon
dispersion from completely above the magnon energy $\omega^{\text m}({\bf q})$ at all ${\bf q}$ deep in the AFM phase to gradually approaching
$\omega^{\text m}({\bf q})$ and eventually dipping below the magnon in parts of the BZ---which happens first at ${\bf q}=(\pi,0)$---as the AFM order
is reduced. For two spinons, with one of them at wavevector ${\bf p}$ and the total wavevector being ${\bf q}$, the bare energy of the spinon pair is then,
\begin{eqnarray}
\tilde\omega^\text{s}({\bf q,p})&=& 2\Delta +c^{\text s}\sqrt{1-\cos^2(p_x)\cos^2(p_y)} \label{s-dispersion}  \\ 
                                &+&c^{\text s}\sqrt{1-\cos^2(q_x-p_x)\cos^2(q_y-p_y)}. \nonumber
\end{eqnarray}
Here it should be noted that, in the simple picture of spinons in the basis of bipartite valence bonds, an $S=1$ excitation corresponds to breaking
a valence bond (singlet), thereby creating a triplet of two spins, one in each of the sublattice A and B \cite{tang13}. The unpaired spins are always confined
to their respective sublattices. There are also two species of magnons, and creating one of them corresponds to a change in magnetization by $\Delta S^z= 1$
or $\Delta S^z= -1$, depending on the sublattice. Since $S^z$ must be conserved, we only need to consider one species of the magnons (e.g, $\Delta S^z= 1$,
which we associate with sublattice A) and that dictates the magnetization of the spinon pair that it can resonate with. 

\begin{figure}[t]
\centering
\includegraphics[width=70mm]{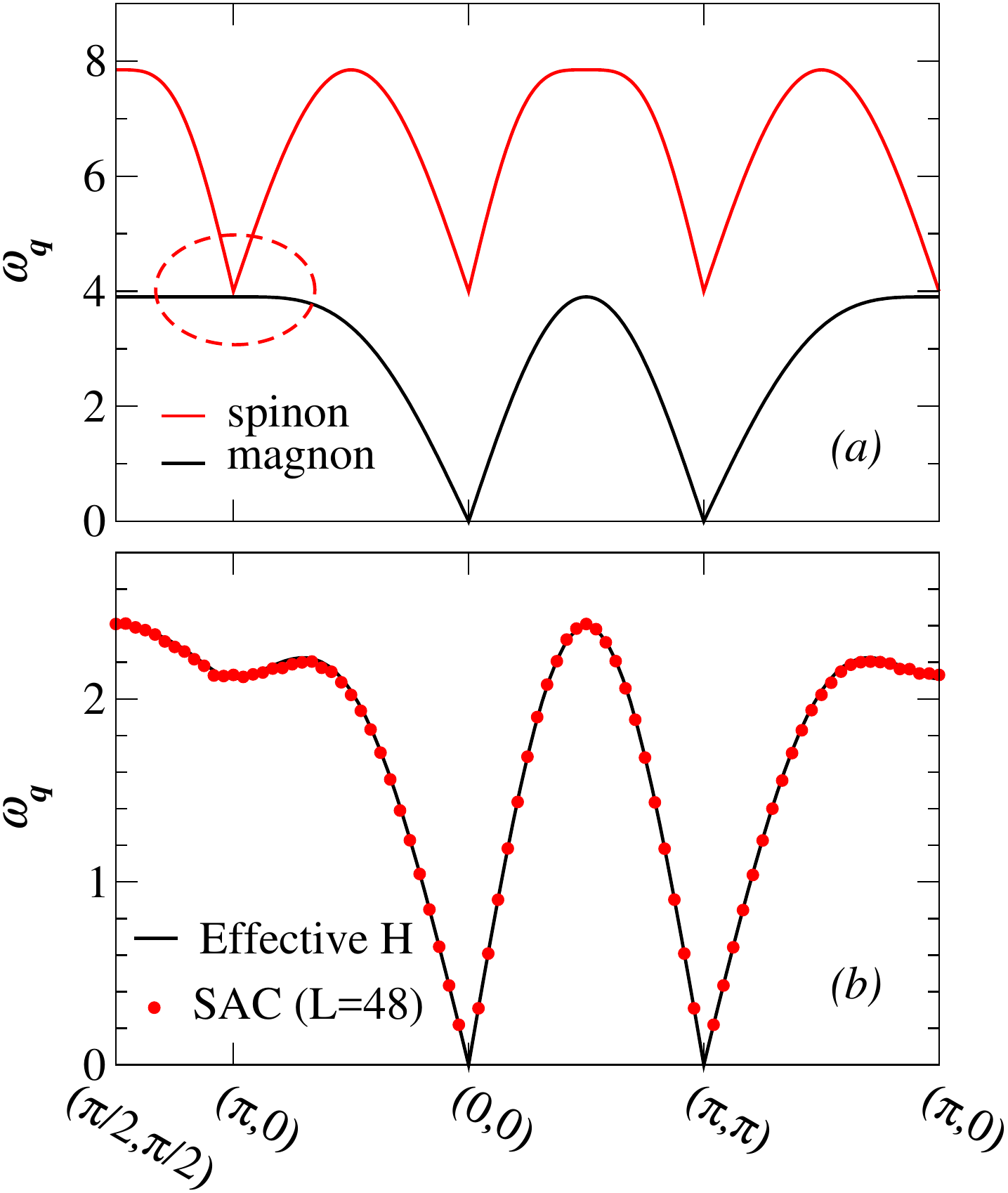}
\caption{(a) Dispersions of the bare excitations of the effective model along a path through the BZ. The lower branch is for the magnon,
  and the upper branch is for a single spinon. The latter is also the lower edge of the two-spinon continuum. In this example, the
  spinons in the circled region close to ${\bf q}=(\pi,0)$ almost touch the magnon band, leading to significant spinon-magnon mixing.
  (b) The black curve shows the lowest energy of the mixed spinon-magnon system obtained with the dispersions in (a)
  and strength $g=5.1$ of the mixing term. The red circles show the results of the QMC-SAC calculations for the Heisenberg model on the
  $L=48$ lattice from Sec.~\ref{sec:hberg}.}
\label{eff-2}
\end{figure}

Instead of adding twice the gap as we do in Eq.~(\ref{s-dispersion}), we could include $\Delta^2$ under each of the square-roots. This
would cause some rounding of the V-shapes of the spinon dispersion. We have confirmed that there are no significant differences between
the two ways of lifting the spinon energies in the coupled spinon-magnon system.

Using second-quantized notation, the non-interacting effective Hamiltonian in the space spanning single-magnon and spinon pair
excitations can be written as 
\begin{eqnarray}
&&H^{\text{eff}-0}_A=\sum_{{\bf q}}\omega^\text{m}({\bf q})d^\dag_{A,{\bf q}}d_{A,{\bf q}}\nonumber\\
&&+\sum_{\bf q,p}\tilde\omega^\text{s}({\bf q,p})c^\dag_{A,{\bf p}}c^\dag_{B,{\bf q-p}}c_{A,{\bf p}}c_{B,{\bf q-p}}
\label{eff-H0}
\end{eqnarray}
where $c^\dag$ ($c)$ and $d^\dag$ ($d$) are the spinon and magnon creation (annihilation) operators, respectively, and there is also an implicit constraint
on the Hilbert space to states with either a single magnon (here on the A sublattice) or two spinons (one on each sublattice). Note that both kinds of
particles are bosons based on the broken-valence-bond picture of the spinons \cite{tang13}.  For brevity of the notation we will hereafter drop the
sublattice index, but in the calculations we always treat the two spinons as distinguishable particles.

Fig.~\ref{eff-2}(a) shows an example of the spinon and magnon dispersions corresponding to the situation we posit for the Heisenberg model. Here the
spinon offset $\Delta$ is sufficiently large to push the entire two-spinon continuum (of which we only show the lower edge) up above the magnon energy,
but at $(\pi,0)$ the spinons almost touch the magnon band. It is clear that any resonance process between the magnon and spinon Hilbert spaces will be
most effective at this point, thus reducing the energy and accounting for the dip in the dispersion found in the QMC study of the Heisenberg model. In
Fig.~\ref{eff-2}(b) we show how well the dispersion relation can be reproduced by the effective model, using a simple spinon-magnon mixing term that we
will specify next.

\begin{figure}[t]
\centering
\includegraphics[width=80mm]{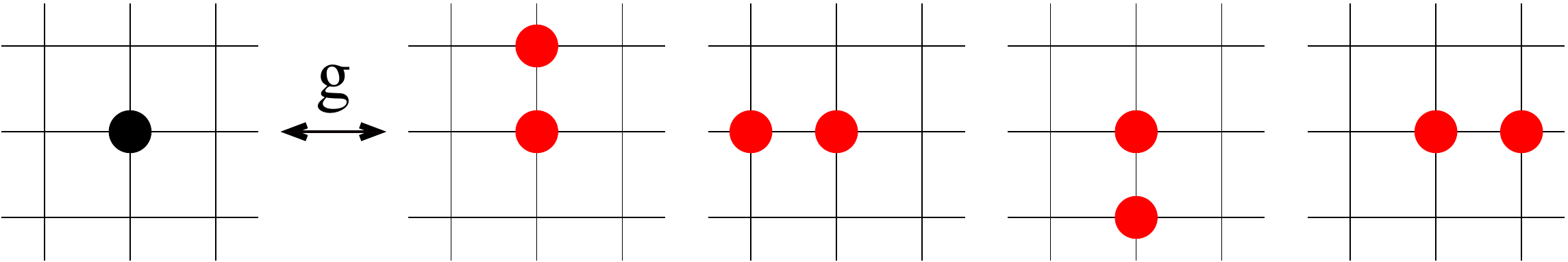}
\caption{Illustration of the mixing process between the magnon (black circle) and the spinon pair (red circles). With mixing strength $g$,
a magnon on a given sublattice splits up into a spinon pair occupying nearest-neighbor sites. The spinon pair can recombine and form a magnon
on the original sublattice.}
\label{eff-1}
\end{figure}

Our basic premise is that the magnon and spinon subspaces mix, through processes where a magnon is split into two spinons and vice versa.
We use the simplest form of this mechanism, where the two spinons are created on neighboring sites, one of those sites being the one on
which the magnon is destroyed. The interaction Hamiltonian in real space is
\begin{equation}   
H^\text I=g\sum_{\bf r,\bf{e}}(c^\dag_{\bf r+\bf{e}}c^\dag_{\bf r}d_{\bf r}+d_{\bf r}^\dag c_{\bf r}c_{{\bf r}+\bf{e}}),
\label{H_I}
\end{equation}
where ${\bf e}$ denotes the four unit lattice vectors as illustrated in Fig.~\ref{eff-1}. In motivating this interaction, we have in mind
how an $S=1$ excitation is created locally, e.g., in a neutron scattering experiment, by flipping a single spin. Spinwave theory describes
the eigenstates of such excitations in momentum space and this leads to the bare magnon dispersion. A spinon in one dimension can be regarded
as a point-like domain wall, and as such is associated with a lattice link instead of a site. However, in the valence bond
basis, the spinons arise from broken bonds and are associated with sites (in any number of dimensions) \cite{tang13}. In this basis, the
initial creation of the magnon also corresponds to creating two unpared spins, and the distinction between a magnon and two deconfined
spinons only becomes clear when examining the nature of the eigenstates (where the spinons may or may not be well-defined particles,
and they can be confined or deconfined). In the actual spin system, the magnon and spinons in the sense proposed here would
never exist as independent particles (not even in any known limit), but the simplified coupled system can still provide
a good description of the true excitations at the phenomenological level, as was also pointed out in the proposal of the AF* state (which also hosts 
topological order that is not present within our proposal) \cite{balents99}. Our
way of coupling the two idealized bare systems according to Eq.~(\ref{H_I}) is intended as a simplest, local description of the mixing of
the two posited parts of the Hilbert space. In the end, beyond its compelling physical picture with key ingredients taken from deconfined quantum 
criticality and the AF* state, the justification of the effective model will come from its ability to reproduce the key properties of the 
excitations of the Heisenberg model.

The magnon-spinon coupling in reciprocal space is
\begin{equation}
H^\text I=\sum_{{\bf q},{\bf p}}I({\bf p})(c_{\bf p}^\dag c^\dag_{{\bf q}-{\bf p}}d_{\bf q}+{\rm h.c.}),
\label{H_I_2}
\end{equation}
where ${\bf q}$ again is the conserved total momentum and ${\bf p}$ is the momentum of the $A$ spinon (more precisely, the above spinon pair creation
operator is $c_{A,\bf p}^\dag c^\dag_{B,{\bf q}-{\bf p}}$), and the form factor corresponding to the
mixing strength $g$ in real space is
\begin{equation}
I({\bf p})=g\sqrt{\frac{2}{N}}\left [\cos (p_x)+\cos (p_y) \right ].
\label{ifunc}
\end{equation}
If this interaction is used directly in a Hamiltonian with the bare magnon and spinon dispersions, we encounter the problem that the ground
state is unstable---the mixing term will push the energy of the lowest excitations below that of the vacuum because the magnon mixes with the
spinon and reduces its energy also at the gapless points.  This behavior is analogous to what would happen to the exciton-polariton spectrum by including
the light-matter interaction without the diamagnetic term. In reality, since $p^2 \to (\mathbf{p}-q{\bf A} )^2$, the minimal exciton-photon
coupling is also responsible for a modification of the photon Hamiltonian, in a way which preserves the gapless spectrum \cite{hopfield58,mahan}.
Following the analogy between magnons/spinon-pairs and photons/excitons, we consider the coupling to arise from a modified spinon-pair operators
by the following substitution in Eq.~(\ref{eff-H0}): 
\begin{equation}
c^\dag_{{\bf p}}c^\dag_{{\bf q-p}} \to c^\dag_{{\bf p}}c^\dag_{{\bf q-p}}+G({\bf q,p})d^\dag_{{\bf q}},
\label{spinons_modified}
\end{equation}
where the mixing function is given by:
\begin{equation}
G({\bf q,p})=\frac{I({\bf p})}{\tilde\omega^\text{s}({\bf q,p})}.
\label{gdef}
\end{equation}
This substitution generates the following effective magnon-spinon Hamiltonian:
\begin{eqnarray}
&&H^{\text{eff}}=\sum_{{\bf q}}\left(\omega^\text{m}({\bf q})+\sum_{\bf p}\tilde\omega^\text{s}({\bf q,p})G^2({\bf q,p})\right)d^\dag_{{\bf q}}d_{{\bf q}}\\\nonumber
&& +\sum_{{\bf p},{\bf q}}\left[\tilde\omega^\text{s}({\bf q,p})c^\dag_{{\bf p}}c^\dag_{{\bf q-p}}c_{{\bf p}}c_{{\bf q-p}} 
  +I({\bf p})c_{\bf p}^\dag c^\dag_{{\bf q}-{\bf p}}d_{\bf q}+{\rm h.c.}\right],
\label{eff-H0-modified}
\end{eqnarray}
Here we see explicitly how the interaction also affects the magnon dispersion (similar to the effect of the diamagnetic term on the exciton-polariton problem),
so that the dressed magnons acquire a slightly renormalized velocity. This procedure guarantees that the ground state is stable and that the
full spectrum of the coupled system is still gapless.

Some aspects of the observed behaviors in the Heisenberg and $J$-$Q$ models can be better reproduced if we also introduce a spinon-spinon interaction
term $V$, to be specified later. Defining the modified magnon dispersion
\begin{equation}
\tilde\omega^\text{m}({\bf q})=\omega^\text{m}({\bf q})+\sum_{\bf p}\tilde\omega^\text{s}({\bf q,p})G^2({\bf q,p}),
\end{equation}
the Hamiltonian in the sector of given total momentum ${\bf q}$ can be written as
\begin{eqnarray}
&&H^{\text{eff}}({\bf q})=\tilde\omega^\text{m}({\bf q})d^\dag_{{\bf q}}d_{{\bf q}}
    +\sum_{{\bf k},{\bf p}}V({\bf k},{\bf p})c^\dag_{{\bf k}}c_{{\bf p}}c^\dag_{{\bf q-k}}c_{{\bf q-p}} \nonumber\\
  &&+\sum_{\bf p}\left[\tilde\omega^\text{s}({\bf q,p})c^\dag_{{\bf p}}c^\dag_{{\bf q-p}}c_{{\bf p}}c_{{\bf q-p}}
  +I({\bf p})c_{\bf p}^\dag c^\dag_{{\bf q}-{\bf p}}d_{\bf q}+{\rm h.c.}\right]. \nonumber\\
\label{eff-H}
\end{eqnarray}
Here it should be noted that, if spinon-spinon interactions are present, $V \not=0$, the definition of the function $G$ changes from Eq.~(\ref{gdef}) in
the following simple way: the non-interacting two-spinon energies $\tilde\omega^\text{s}({\bf q,p})$ should be replaced by the eigenenergies of the
interacting 2-spinon subsystem, and the momentum label ${\bf p}$ accordingly changes to a different index labeling the eigenstates. The mixing term
is also transformed accordingly by using the proper basis in Eq.~(\ref{spinons_modified}).

We study the effective Hamiltonian by numerical ED on $L\times L$ lattices with $L$ up to $64$. Our effective model is clearly very
simplified and one should of course not expect it to provide a fully quantitative description of the excitations of the many-body spin Hamiltonians.
Nevertheless, it is interesting that the parameters $c^m,c^s,\Delta,$ and $g$ can be chosen such that an almost perfect agreement with
the Heisenberg magnon dispersion obtained in Sec.~\ref{sec:hberg} is reproduced, as shown in Fig.~\ref{eff-2}(b) (where no spinon-spinon interactions are
included). In the following we will not attempt to make any further detailed fits to the results for the spin systems, but focus on the general behaviors
of the model and how they can be related to the salient features of the Heisenberg and $J$-$Q$ spectral functions.

\subsection{Mixing states and spectral functions}

\begin{figure}[t]
\centering
\includegraphics[width=70mm]{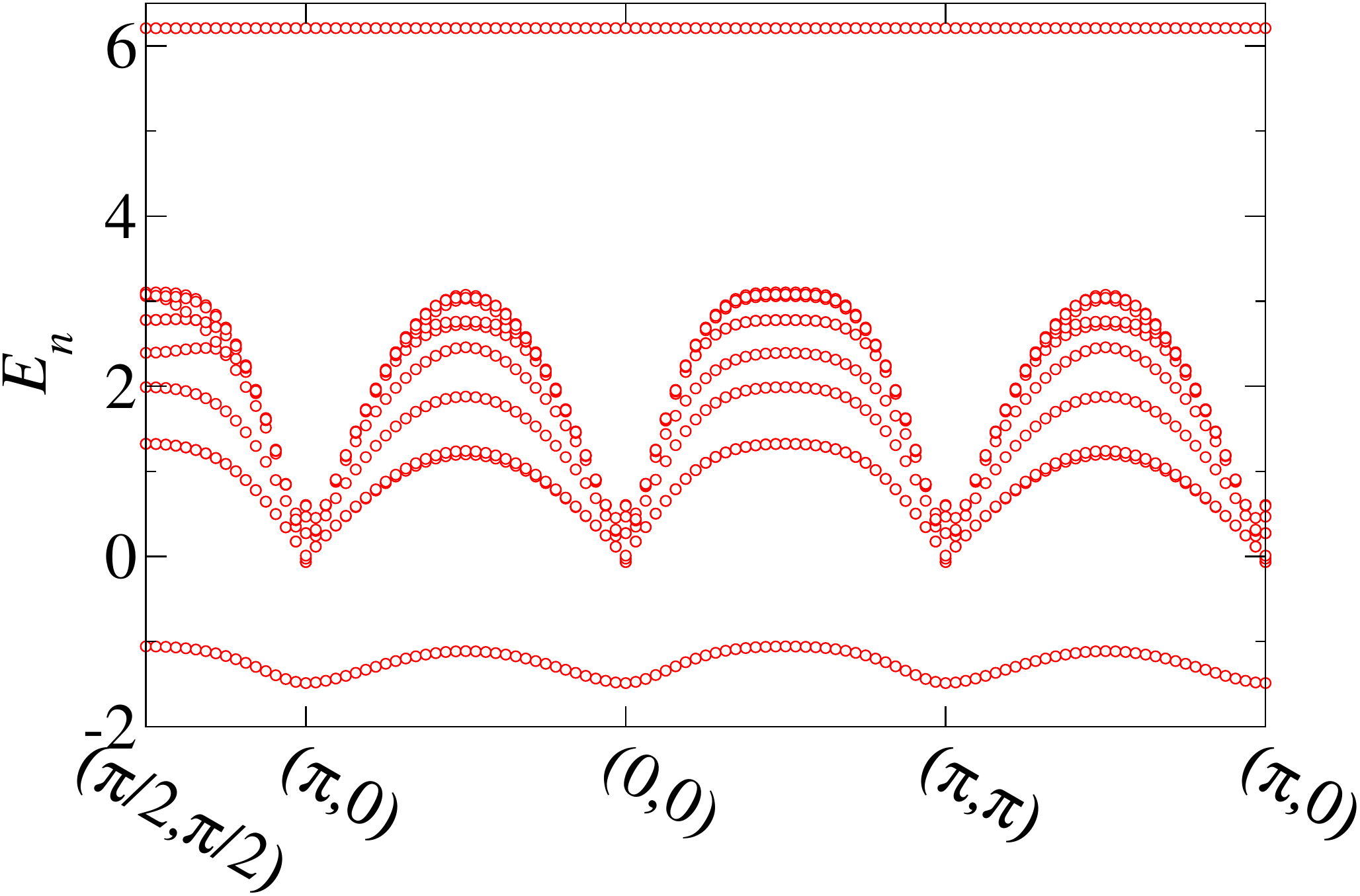}
\caption{Energy levels versus the total wavevector of two spinons interacting through a potential $V(r)=-6.2{\rm e}^{-r/2}$.
  The bare dispersion relation of the single spinon is given by Eq.~(\ref{free-s-dispersion}) with $c^s=3.1$. We only show a few of the
  levels between the lower and higher energy bound.}
\label{levels}
\end{figure}

For a given total momentum ${\bf q}$, the eigenstates $|n,{\bf q}\rangle$ of the effective Hamiltonian in Eq.~(\ref{eff-H}) have overlaps
$\langle n,{\bf q}|{\bf q}\rangle$ with the bare magnon state $|{\bf q}\rangle$. Without spinon-spinon interactions ($V=0$), with the bare spinons
above the magnon band for all ${\bf q}$, and when the mixing parameter $g$ is suitable for describing the Heisenberg model [i.e., giving good agreement
with the QMC dispersion relation, as in Fig.~\ref{eff-2}(b)], we find that all but the first and the last of these overlaps become very small when the
lattice size $L$ increases. Thus, the two particular states are magnon-spinon resonances and the rest are essentially free states of the two-spinons.
When attractive spinon-spinon interactions are included, the picture changes qualitatively, with the magnon also mixing in strongly with all spinon
bound states. An example of spinon levels in the presence of spin-spin interactions are shown in Fig.~\ref{levels}, where a number of bound states
separated by gaps can be distinguished. The stronger mixing with the bound states is simply a reflection of the fact that two bound spinons have
a finite probability to occupy nearest-neighbor sites, so that the mixing process with the magnon (Fig.~\ref{eff-1}) can take place, while the
probability of this vanishes when $L \to \infty$ for free spinons. Note that the total overlap $\langle n,{\bf q}|{\bf q}\rangle$ summed over
all free-spinon states can still be non-zero, due to the increasing number of these states.

The fact that the dispersion relation resulting from $H^{\text{eff}}$ can be made to match the QMC-SAC results for the Heisenberg model (Fig.~\ref{eff-2})
is a tantalizing hint that the dispersion anomaly at ${\bf q}=(\pi,0)$ may be a precursor of spinon deconfinement as some interaction brings the system
further toward the AFM--VBS transition. In the weak magnon-spinon mixing limit, the lowest-energy spinons will, in the absence of attractive spinon-spinon
interactions $V$, deconfine close to ${\bf q}=(\pi,0)$ if the spinon continuum falls below the magnon band at this wave vector, while the magnon-spinon
resonance remains lowest excitation in parts of the BZ where the bare spinons stay above the magnon. The resonance state should still be considered as a magnon,
as the spinons are spatially confined and constitute an internal structure to the magnon.

This simple behavior, which essentially follows from the postulated bare dispersion relations, is very intriguing because it is precisely what we observed
in Sec~\ref{sec:jq} for the $J$-$Q$ model when $Q$ is turned on but is still far away from the deconfined critical point. We found (Figs.~\ref{jq} and
\ref{fss}), that the low-energy magnon pole vanishes at $(\pi,0)$, while it remains
prominent at $(\pi/2,\pi/2)$. Thus, we propose that increasing $Q/J$ corresponds to a reduction of the energy shift $\Delta$ in the bare spinon energy
in Eq.~(\ref{s-dispersion}), reaching $\Delta=0$ at the deconfined quantum-critical point. At the same time the bare magnon and spinon velocities should also
evolve in some way. The observation that the $(\pi/2,\pi/2)$ magnon survives even at the critical point would suggest that the magnon band remains below
the spinon continuum at this wave vector.

Let us now investigate the spectral function of the effective model. Within the model, the spectral function corresponding to the dynamic spin
structure factor of the spin models is that of the magnon creation operator $d^\dagger_{\bf q}$
\begin{equation}
S({\bf q},\omega)=\sum_{n}|\langle n|d^\dagger_{\bf q}|{\rm vac} \rangle|^2\delta(\omega-E_n),
\label{sqweff}
\end{equation}
where $|{\rm vac} \rangle$ is the vacuum representing the ground state of the spin system and $E_n$ is the energy of the eigenstate $|n \rangle$. The
matrix element is nothing but the absolute-squared of the magnon overlap $\langle n,{\bf q}|{\bf q}\rangle$ discussed above. Thus, with non-interacting
spinons the spectral function consists of two $\delta$-functions, corresponding to the two spinon-magnon resonance states, and a weak continuum arising from
a large number of deconfined 2-spinon states. The situation changes if we include spinon-spinon interactions. Then, as mentioned above, the spinon
bound states mix more significantly with the magnon and gives rise to more spectral weight in Eq.~(\ref{sqweff}) away from the edges of the spectrum,
and the $\delta$-function at the upper edge essentially vanishes. To attempt to model the spinon-spinon interactions quantitatively would be beyond
the scope of the simplified effective model, but by considering a reasonable case of short-range interactions we will observe interesting features
that match to a surprisingly high degree with what was observed in the spin systems.

\begin{figure}[t]
\centering
\includegraphics[width=70mm]{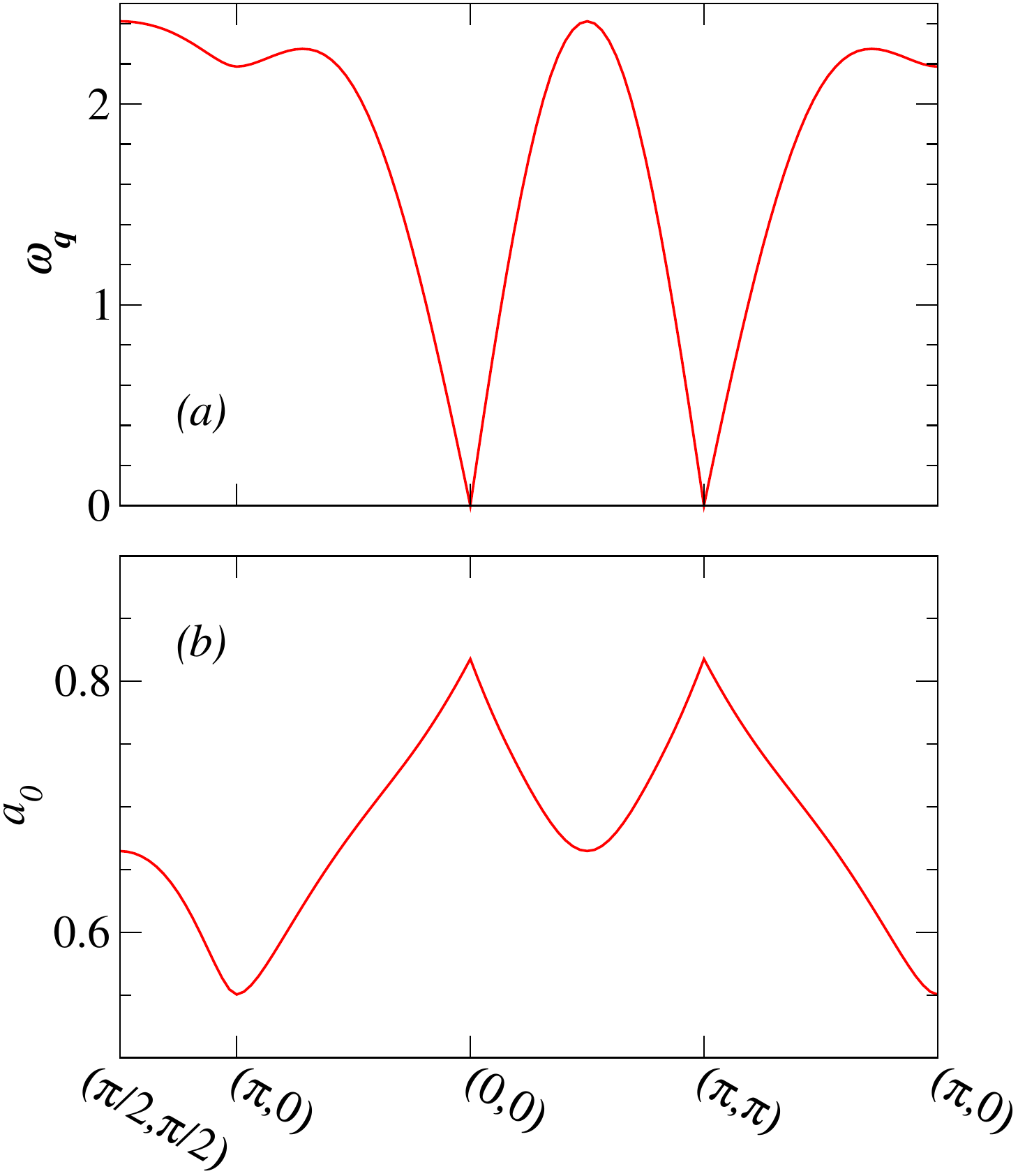}
\caption{Dispersion relation (a) and wavevector dependence of the relative weight of the magnon pole (b) calculated with the effective
Hamiltonian with the parameters $c^m=3.1, c^s=3.1, \Delta=1.94, g=1.86$, and the spinon-spinon potential $V(r)=-6.2{\rm e}^{-r/2}$.}
\label{effresults}
\end{figure}

The ${\bf q}$ dependence of the total spectral weight of the spin system cannot be modeled with our approach here, because the effective model completely
neglects the structure of the ground state, replacing it by trivial vacuum, and the magnon creation operator is also an oversimplification of the spin
operator. Because of these simplifications the total spectral weight is unity for all ${\bf q}$. A main focus in Secs.~\ref{sec:hberg} and \ref{sec:jq}
was on the relative weight $a_0({\bf q})$ of the leading magnon pole, and this quantity does have its counterpart in Eq.~(\ref{sqweff});
\begin{equation}
a_0({\bf q})=|\langle n=0 | d^\dagger_{\bf q}|{\rm vac} \rangle|^2 = |\langle 0|{\bf q}\rangle|^2,
\end{equation}
where $|n=0\rangle\rangle$ is the lowest-energy eigenstate and $a_0({\bf q})$ can be compared with the QMC/SAC results in Fig.~\ref{haf-5}. Given that the
Hilbert space of the effective model contains only a single magnon, the spectral function should correspond to the transverse component in situations where
the transverse and longitudinal contributions are separated (e.g., polarized neutron scattering). 

We now include attractive spinon-spinon interactions such that bare (before mixing with the magnon) bound states are produced,
as in Fig.~\ref{levels}. The other model parameters are again adjusted such that the dispersion relation resembles that in the Heisenberg model,
with the anomaly at ${\bf q}=(\pi,0)$. The resulting dispersion (location of the dominant $\delta$-function, which constitutes the lower edge of the spectral
function) as well as the relative magnon amplitude are graphed in Fig.~\ref{effresults}. The dispersion relation is very similar to that obtained
without spinon-spinon interactions in Fig.~\ref{eff-2}. Comparing the amplitude $a_0({\bf q})$ in Fig.~\ref{effresults}(b) with the Heisenberg results in Fig.~\ref{haf-5},
we can see very similar features, with minima and maxima at the same wavevectors, though the variations in the amplitude are larger in the
Heisenberg model.

\begin{figure}[t]
\centering
\includegraphics[width=70mm]{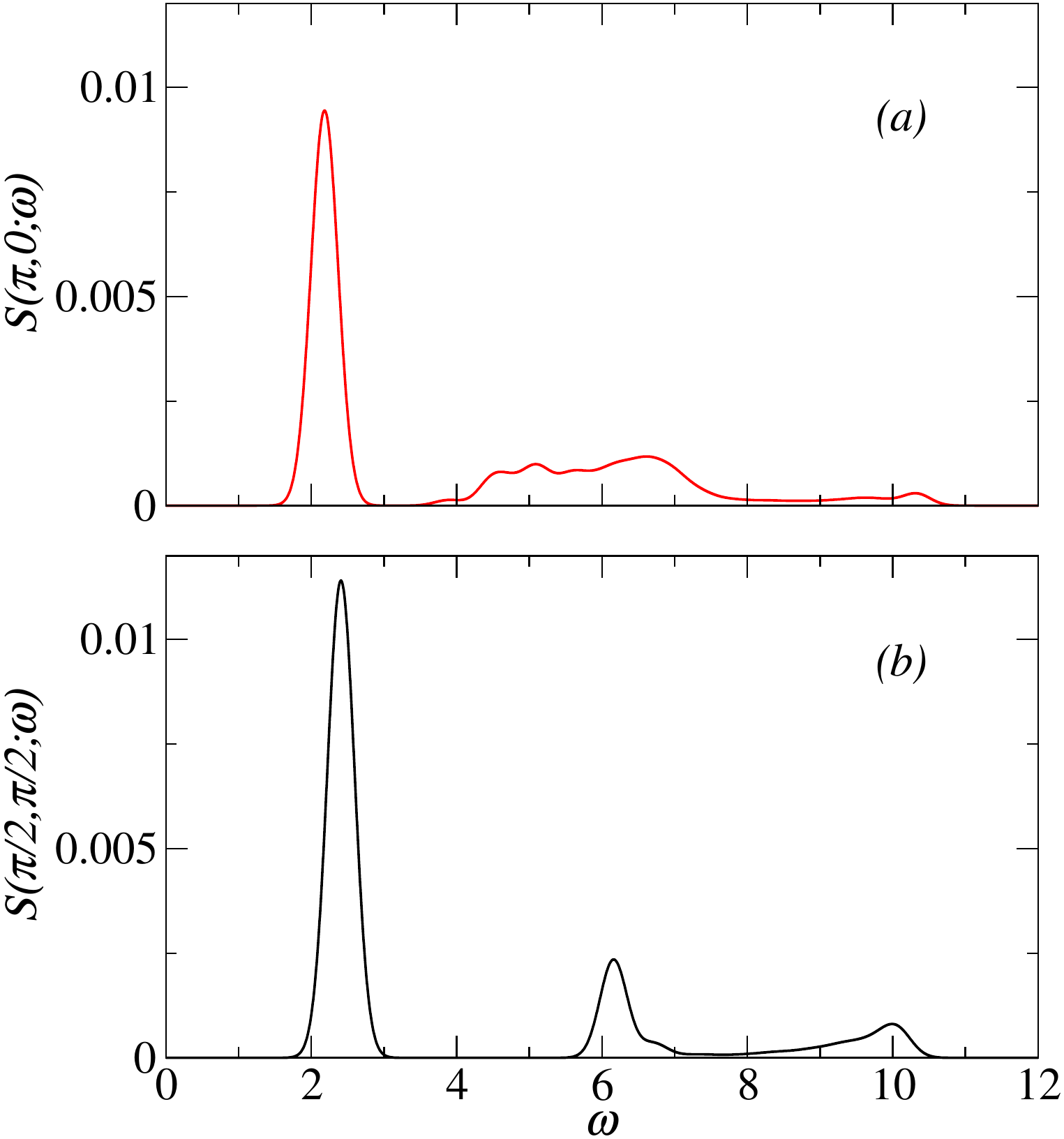}
\caption{Spectral functions of the effective model at (a) ${\bf q}=(\pi,0)$ and (b) ${\bf q}=(\pi/2,\pi/2)$, using model parameters corresponding
  to the Heisenberg model; $c^m=3.1, c^s=3.1, \Delta=1.94, g=1.86$ and the spinon-spinon potential $V(r)=-6.2{\rm e}^{-r/2}$ (same as used
  in Fig.~\ref{levels} and \ref{effresults}). The $\delta$-functions in the exact spectral function (computed here using
  an $L=64$ lattice) have been broadened for visualization.}
\label{hbsqweff}
\end{figure}

\begin{figure}[t]
\centering
\includegraphics[width=70mm]{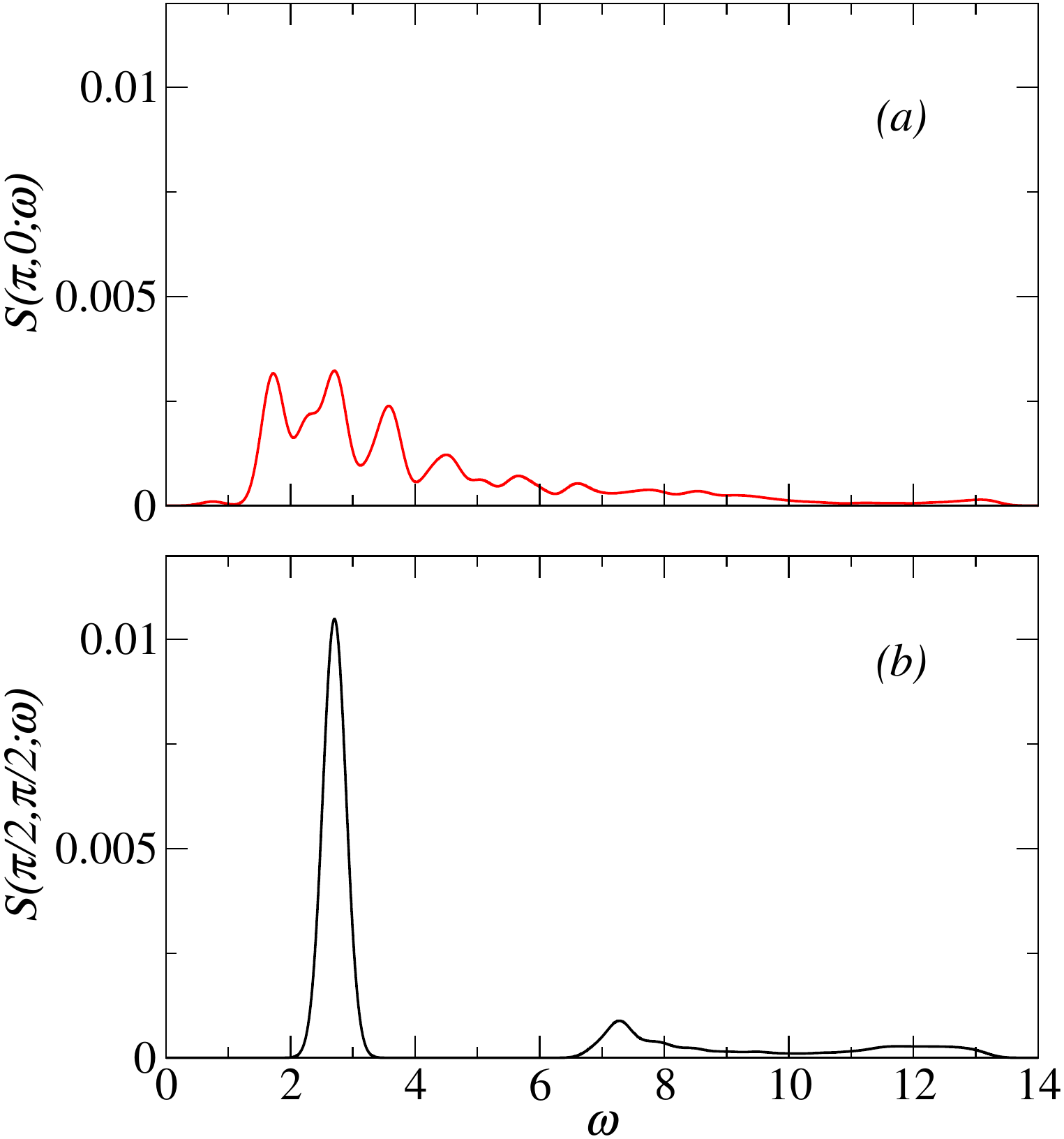}
\caption{Spectral functions as in Fig.~\ref{hbsqweff}, but with the parameters of the effective model chosen to give behaviors
similar to the $J$-$Q$ model with $Q \approx J$; $c^m=3.1, c^s=6.2, \Delta=0.39, g=1.86$ and the spinon-spinon potential 
$V(r)=-6.2{\rm e}^{-r/2}$.}
\label{jqsqweff}
\end{figure}

The full spectral functions at ${\bf q}=(\pi,0)$ and $(\pi/2,\pi/2)$ are displayed in Fig.~\ref{hbsqweff}. Here we have broadened all $\delta$-functions
to obtain continuous spectral functions. As already discussed, the prominent $\delta$-function corresponding to the magnon is similar to what is observed
in the Heisenberg model, though clearly the shapes of the continua above the main $\delta$-function are different from those in Fig.~\ref{haf-2}. Upon
reducing the spinon energy offset $\Delta$ so that the bare energy falls below the magnon energy close to ${\bf q}=(\pi,0)$, we observe a very interesting
behavior in Fig.~\ref{jqsqweff}. We see that the main magnon peak is washed out, due to decay into the lower spinon states. This is very similar to what we
found for the $J$-$Q$ model in Sec.~\ref{sec:jq}, where already a relatively small value of $Q/J$ led to a broad spectrum without magnon pole at ${\bf q}=(\pi,0)$.
At $(\pi/2,\pi/2)$ the magnon pole remained strong, however, and this is also what we see for the effective model in Fig.~\ref{jqsqweff}. Without
spinon-spinon interactions, when the bare magnon is inside the spinon continuum a sharp (single $\delta$-function) spinon-magnon resonance remains in
inside the continuum of free spinon states. Thus, for the magnon pole to completely decay, spinon-spinon interactions are essential in the
effective model.

These results for a simple effective model provide compelling evidence for the mechanism of magnon-spinon mixing outlined above. The results
also suggest that the absence of magnon pole at and close to ${\bf q} = (\pi,0)$ does not necessarily imply complete spinon deconfinement, as we
have to include explicitly attractive interactions in the effective model in order to reproduce the behavior in the full spin systems. Weak
attractive spinon-spinon interactions have previously been detected explicitly in the $J$-$Q$ model at the deconfined critical point \cite{tang13},
and they are also expected based on the field-theory description, where the spinons are never completely deconfined due to their coupling to an
emergent gauge field \cite{senthil04a}. The loss of the magnon pole observed here then signifies that the magnon changes character, from a single
spatially well-resolved small resonance particle to a more extended particle (with more spinon characteristics) as a weak $Q$ interaction is turned on, 
and finally the particle completely disintegrating into a continuum of weakly bound spinon pairs and deconfined spinons.

\section{Conclusions}
\label{sec:summary}

We have investigated the long-standing problem of the excitation anomaly at wavevectors ${\bf q} \approx (\pi,0)$ in the spin-$1/2$ square lattice
Heisenberg antiferromagnet, and established its relationship to deconfined quantum criticality by also studying the $J$-$Q$ model. Using an improved
stochastic (sampling) method for analytic continuation of QMC correlation functions, we have been able to quantify the evolution of the magnon pole
in the dynamic structure factor $S({\bf q},\omega)$ as the AFM order is weakened with increasing ratio $Q/J$, all the way from the Heisenberg limit
$(Q=0$) to the deconfined critical point at $Q/J \approx 22$. For the Heisenberg model, our results agree with other numerical approaches (series
expansions \cite{zheng05} and continuous similarity transformations within the Dyson-Maleev formalism \cite{Powalski15}) 
and also with recent inelastic neutron scattering experiments of
the quasi-2D antiferromagnet CFTD \cite{piazza15}. Upon increasing $Q/J$, we found a rapid loss of single-magnon weight at ${\bf q} \approx (\pi,0)$,
but not at ${\bf q} \approx (\pi/2,\pi/2)$, where the magnon pole remains robust even at the critical point. At first sight these behaviors appear
surprising, but we can consistently explain them through the proposed connection to deconfined quantum criticality.

Motivated by the numerical results, we have constructed an effective model of magnon-spinon mixing that can phenomenologically explain not only the
fragile, almost fractionalized $(\pi,0)$ magnon of the Heisenberg model and its decay into spinon pairs with increasing $Q/J$, but also establishes the reason
of the stability of the $(\pi/2,\pi/2)$ magnon in the $J$-$Q$ model for large $Q$ (as discovered with the QMC-SAC calculations). The essential
ingredient is a gapped spinon band with a dispersion minimum at $(\pi,0)$, for which we find motivation in the fact that this point becomes gapless at the
deconfined quantum critical point. If the continuum of bare spinon excitations remains above the
magnon band throughout the BZ (as in Fig.~\ref{eff-2}), then the lowest excitations are always magnons. However, since the two bands are coupled
in the effective model, via a term that destroys a magnon and creates two spinons (as well as its conjugate destroying the spinons and creating a magnon),
the magnons fluctuate in and out of the spinon space, and this effect is the largest at the point in the BZ where the gap between the two bare branches
is the smallest, i.e., at ${\bf q}=(\pi,0)$. We find that this effect can account quantitatively for the dip in the magnon dispersion relation,
and qualitatively the wavevector dependence of the relative weight of the $\delta$-function at the lower edge of the spectrum is also captured.

Within this effective model, the deconfinement mechanism in the $J$-$Q$ model is explained as the bare spinon dispersion dipping below the magnon at
${\bf q}=(\pi,0)$. This can happen already for small $Q/J$, far away from the AFM--VBS transition, because the bare magnon-spinon gap is already small for $Q=0$.
As $Q/J$ increases, an increasing fraction of the BZ becomes deconfined, until finally the gapless spinons deconfine at the critical point. Our QMC-SAC
results indicate that the excitations at higher energy remain confined, as exemplified by ${\bf q}=(\pi/2,\pi/2)$. Within the effective model this
follows from the bare spinon dispersion staying above the magnon band in this region of wavevectors.

Clearly the effective model should not be taken as a quantitative description of the Heisenberg and $J$-$Q$ systems; motivated by aspects of deconfined
quantum-criticality and the AF* state, we have introduced it mainly as a phenomenological tool for elucidating the behaviors observed in the QMC studies
of the model Hamiltonians. Nevertheless, it is remarkable how well the essential observed features are captured and how otherwise non-intuitive aspects of the 
deconfinement mechanism follow naturally from the magnon-spinon mixing under mild assumptions on the bare parameters of the effective model. Thus, 
even in the absence of a strict microscopic derivation, the effective model can be justified by its many non-trivial confirmed predictions.

Considering the mechanism leading to the loss of magnon pole with increasing $Q$, it is interesting to note that it does not appear to involve
significant broadening of the $\delta$-function, but instead the spectral weight of this peak is distributed out into the continuum by the spinon
mixing process. This is in accord with the general belief that quantum antiferromagnets with collinear order lack the damping processes that
cause the broadening of the magnon pole in frustrated, non-collinear magnets \cite{chernyshev06,ma16,kamiya17}. Our proposed mechanism of spinon mixing
is, thus, very different from standard magnon damping.

The scenario of a nearly fractionalized magnon in the Heisenberg model does not necessarily stand in conflict with the expansion in multi-magnon processes
\cite{Powalski15,Powalski17}, which can account for the dynamic structure factor without invoking any spinon mixing effects. We have only discussed the 
effective model of the excitations at the level of a single magnon and its mixing with the spinon continuum, and our results for the Heisenberg model show that 
the magnon is significantly dressed by spinons around ${\bf q}=(\pi,0)$ but is not yet fractionalized. The magnon-spinon mixing then represents a description
of the internal structure of the magnon, and we have not considered the further effects of multi-magnon processes. It is remarkable that the results of 
Ref.~\onlinecite{Powalski17} match the experimental data (and also numerical data for the Heisenberg model) so well without taking into account the internal 
spinon structure of the magnons, if indeed this structure is present. Here we can draw a loose analogy with nuclear physics, where the 
inter-nucleon force has an effective description in terms of exchange of mesons (pions) between nucleons. Yukawa proposed mesons as the carriers of the 
force without knowledge of the quark structure of the nucleons and mesons that is ultimately involved in the interaction (residual strong force) process, 
and quantitatively satisfactory results in nuclear physics are obtained with the effective interaction (and calculations with the full strong force
between quars mediated by gluons are in practice too complicated to work with quantitatively). The significant attractive interaction between magnons in 
the Heisenberg model \cite{Powalski15,Powalski17} might perhaps similarily be regarded as mediated by spinon pairs (which themselves constitute magnons), 
and, by the pion analogy, the magnons and their residual attractive interactions could also provide an accurate description of the excitations without 
invocing the internal spinon structure. To investigate the relationship between the two pictures further, it would be interesting to treat the $J$-$Q$ 
model with the method of Ref.~\onlinecite{Powalski17}. Based on our scenario we predict that the multi-magnon expansion should break down rapidly close 
to ${\bf q}=(\pi,0)$as the $Q$ interaction is turned on but remain convergent at low energies until the system comes close to the deconfined 
quantum-critical point.

The fragility of the magnons at and close to ${\bf q}=(\pi,0)$ suggests that these excitations may become completely fractionalized also by other interactions
than the $Q$-terms considered here, e.g., ring exchange or longer-range pair exchange. These interactions have recently also been investigated
in the context of possible topological order and spinon excitations in the cuprates \cite{chatterjee17}. Earlier the so-called AF* state had
been proposed, largely on phenomenological grounds, where topological order coexists with AFM order and there is a spinon continuum similar 
to the one in our effective model \cite{balents99,senthil00}. Though in our scenario the reason for the spinon continuum is different---the 
proximity to a deconfined quantum critical point---a generic conclusion valid in either case is that spinon deconfinement can set in at 
${\bf q}=(\pi,0)$ well before any ground state transition at which the low-energy spinons deconfine.

In this context the quasi-2D square-lattice antiferromagnet Cu(pz)$_2$(ClO$_4$)$_2$ is very interesting. It has a weak frustrated next-nearest-neighbor 
coupling and has been modeled within the $J_1$-$J_2$ Heisenberg model \cite{tsyrulin09}. Neutron scattering experiments on the material and 
series-expansion calculations for the model show an even larger suppression of the $(\pi,0)$ energy than in the pure Heisenberg model, similar to what 
we have observed in the presence of a weak $Q$ interaction. The experimental $(\pi,0)$ line shape also seems to have a smaller magnon pole than CFTD, 
in accord with our scenario of a fragile magnon pole, although we are not aware of any quantitative analysis of the weight of the magnon pole and
no line-shape calculations were reported in Ref.~\onlinecite{tsyrulin09}. It would clearly be intersting to carry out neutron experiments at higher resolution 
and to make detailed comparisons with calculations beyond the dispersion relation. 

Ultimately the $J_1$-$J_2$ system should be different from the $J$-$Q$ model, because the deconfined quantum critical point of the latter most likely is replaced 
by an extended gapless spin liquid phase of the former \cite{hu13,gong14,morita15,wang17}. 
However, since this phase should also be associated with deconfined spinons, the 
evolution of the excitations as this phase is approached may be very similar to what we have discussed within the $J$-$Q$ model on its approach to the 
deconfined quantum critical point. A state with topological order and spinon excitations may instead be approached when strong ring-exchange interactions are added \cite{chatterjee17}, but
given that $J$ is weak in Cu(pz)$_2$(ClO$_4$)$_2$ these interactions may not play a significant role in this case. Ring exchange should be more important
in Sr$_2$CuO$_2$Cl$_2$, where excitation anomalies have also been observed \cite{guarise10}.

The magnetic-field ($h$) dependence of the excitation spectrum of Cu(pz)$_2$(ClO$_4$)$_2$ was also studied in Ref.~\onlinecite{tsyrulin09}. Since the energy scale
of the Heisenberg exchange is even smaller than in CFTD, it was possible to study field strengths of order $J$ and observe significant changes in the
dispersion relation and the $(\pi,0)$ line shape. The methods we have developed here can also be applied to systems in an external magnetic field 
and it would be interesting to study the dynamics of the $J$-$Q$-$h$ model. Some results indicating destabilization of magnons due to the field in 
the Heisenberg model are already available \cite{syljuasen08b}, and our improved analytic continuation technique could potentially improve on the 
frequency resolution.

\begin{acknowledgments}
We thank Wenan Guo, Akiko Masaki-Kato, Andrey Mishchenko, Martin Mourigal, Henrik R{\o}nnow, Kai Schmidt, Cenke Xu, and Seiji Yunoki for useful 
discussions. Experimental data from  Ref.~\cite{piazza15} were kindly provided by N. B. Christensen and H. M. R{\o}nnow. H.S. was supported by 
the China Postdoctoral Science Foundation under Grant Nos.~2016M600034 and 2017T100031. St.C was funded by the NSFC under Grant Nos.~11574025 and U1530401. 
Y.Q.Q. and Z.Y.M. acknowledge funding from the Ministry of Science and Technology of China through the National Key Research and Development Program
under Grant No.~2016YFA0300502, and from the NSFC under Grant Nos.~11574359 and 11674370, as well as the National Thousand-Young Talents Program of China.
A.W.S. was funded by the NSF under Grant Nos.~DMR-1410126 and DMR-1710170, and by the Simons Foundation. In addition H.S., Y.Q.Q., and Sy.C. thank 
Boston University's Condensed Matter Theory Visitors program for support, and A.W.S. thanks the Beijing CSRC and the Institute of Physics, Chinese 
Academy of Sciences for visitor support. We thank the Center for Quantum Simulation Sciences at the Institute of Physics, Chinese Academy of Sciences, 
the Tianhe-1A platform at the National Supercomputer Center in Tianjin, and Boston University's Shared Computing Cluster for their technical support and
generous allocation of CPU time.
\end{acknowledgments}

\appendix

\section{Covariance in QMC and synthetic data}
\label{app:covar}

As discussed in Sec.~\ref{sec:sac1} the QMC-computed imaginary time data $\bar G(\tau_i)$ for different $i$ are
correlated, and it is well known \cite{jarrell96} that this has to be taken into account in any statistically proper
analytic continuation procedure (though in practice good results can still be obtained with just the diagonal elements $\sigma_i$,
if they are sufficiently small). While the covariance may seem like a nuisance, there is actually a silver lining, in that
correlations between different $\tau$-points typically imply that the data are actually better than the individual
statistical errors $\sigma_i$ might indicate.

As an extreme example of the above, imagine a situation in which all data points are perfectly correlated in the sense that
the computed $\bar G_i$ (over a bin or the whole simulation) is of the form
\begin{equation}
G_i = G^{\rm exact}_i (1+\sigma)
\end{equation}
for all $i$, where $\sigma$ is the common noise source. Then, upon normalization, $G_i \to G_i/G_0$, one obtains the exact
value $G^{\rm exact}_i/G^{\rm exact}_0$ (where the subscript $0$ corresponds to $\tau=0$). In reality the noises for different $\tau$-points
are not perfectly correlated, but have an autocorrelation function that decays with $\tau$, but nevertheless the presence of
covariance corresponds to additional information content in the data set, and this information can improve the frequency
resolution when compared to the case of no off-diagonal elements of $C$ and the same values of all $\sigma_i=C_{ii}$. Here
we show some examples of covariance-effects in QMC data, and also explain how we build in correlated noise in synthetic data.

\subsection{Real QMC data}

\begin{figure}[t]
\centering
\includegraphics[width=65mm]{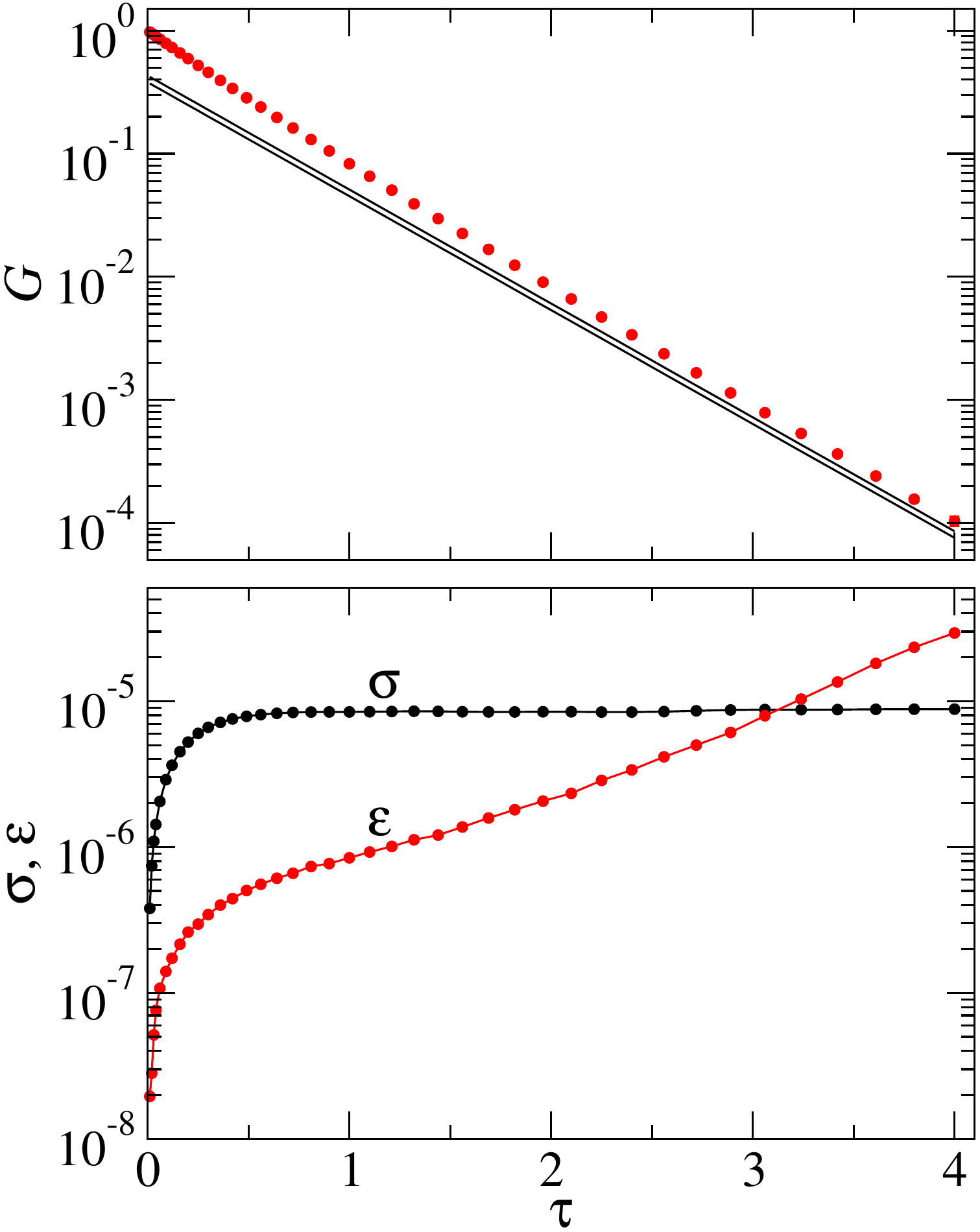}
\caption{(a) Imaginary-time correlation function at $q=(\pi,0)$ for 2D Heisenberg lattice with $L=48$, computed in SSE QMC
  simulations at $\beta=192$ (giving $T=0$ results for all practical purposes). The two straight lines correspond to the
  contribution for the leading $\delta$-function obtained in the SAC procedure, with amplitude $a_0=0.405 \pm 0.025$. (b)
  The statistical errors (diagonal elements of the covariance matrix) $\sigma(\tau)$ and the eigenvalues of the covariance
  matrix (ordered from smallest to largest).}
\label{greal}
\end{figure}

In Fig.~\ref{greal} we show an example of data underlying the SAC calculations in Sec.~\ref{sec:hberg}; at the most
interesting wavevector, ${\bf q}=(\pi,0)$, for a system with $L=48$. We have here used a quadratic $\tau$-grid, in order to take
advantage of the reduced error bars close to $\tau=0$ after normalizing to $G(0)=1$, while not including an excessively large
number of points (in which case there is a lot of redundancy in the correlated data and it also becomes difficult to diagonalize
the covariance matrix). We only include data points for which the relative errors $\sigma_i/\bar G_i$ are less than $10\%$.

Fig.~\ref{greal}(a) shows the $G(\tau)$ data on a lin-log scale, so that a pure exponential decay (arising from a spectrum
with a single $\delta$-function) corresponds to a straight line. From the analysis in Sec.~\ref{sec:hberg} we have that the amplitude
of the magnon $\delta$-function is $a_0 = 0.405 \pm 0.025$ and its frequency is $\omega_0 \approx 2.13$. The two straight lines in the
figure correspond to the contribution from this $\delta$-function when the amplitude is the mean value plus or minus one error bar,
i.e., $0.38$ and $0.43$, respectively. These lines are still significantly below the data points and it is also clear that the data
have not quite converged to a pure straight line at the largest $\tau$ available. Therefore, it is not easy to extract $a_0$ and
$\omega_0$ from a simple exponential fit to the large-$\tau$ data, and the SAC procedure with the special treatment of the magnon
pole should be an optimal way to take into account the effects of the continuum.

\begin{figure}[t]
\centering
\includegraphics[width=70mm]{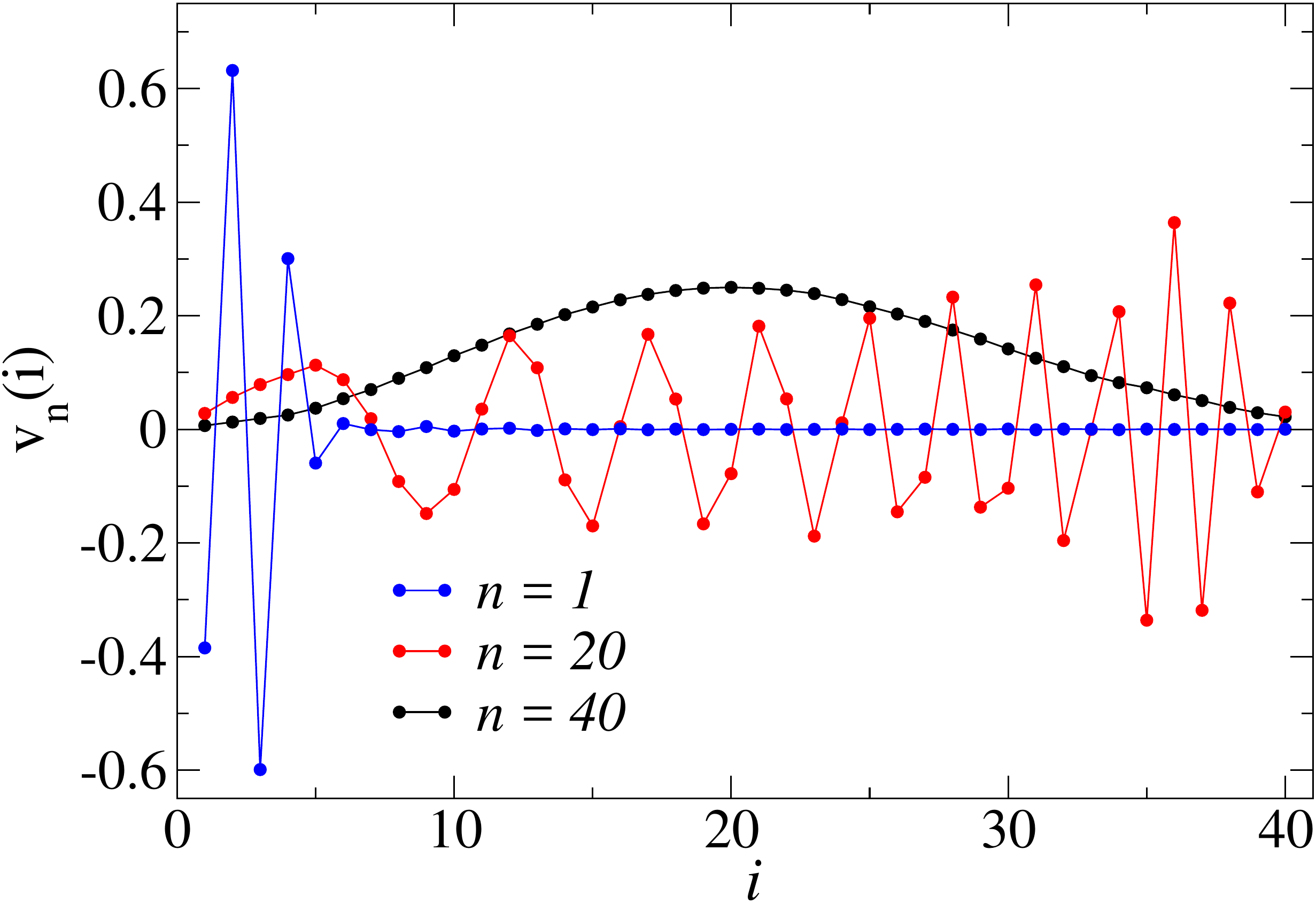}
\caption{The eigenvectors corresponding to the smallest ($n=1$) and largest ($n=40$) eigenvalues of the covariance matrix
in Fig.~\ref{greal}, as well as one from the middle of the eigenvalue spectrum ($n=20$).}
\label{vectors}
\end{figure}

It is also interesting to examine the eigenvectors of the covariance, i.e., the linear combinations,
\begin{equation}
{\bf v_n} = \sum_i v_n(i)G(\tau_i),
\end{equation}
of the imaginary-time data points that fluctuate independently of each other in the QMC simulations. Figure~\ref{vectors} shows three of the
normalized eigenvectors corresponding to the eigenvalues in Fig.~\ref{greal}. Note that the normalization of $G(\tau_i)$ has already removed a
significant component of the covariance---the uniformly fluctuating component---and without the normalization the largest eigenvector
has the most weight for small $\tau_i$, instead of being shifted to higher $\tau_i$ with the normalized data set (seen for $n=1$ in the
figure). The vector corresponding to smallest eigenvalue has alternating positive and negative values and decays rapidly with
$\tau_i$.

\subsection{Synthetic data}

\begin{figure}[t]
\centering
\includegraphics[width=65mm]{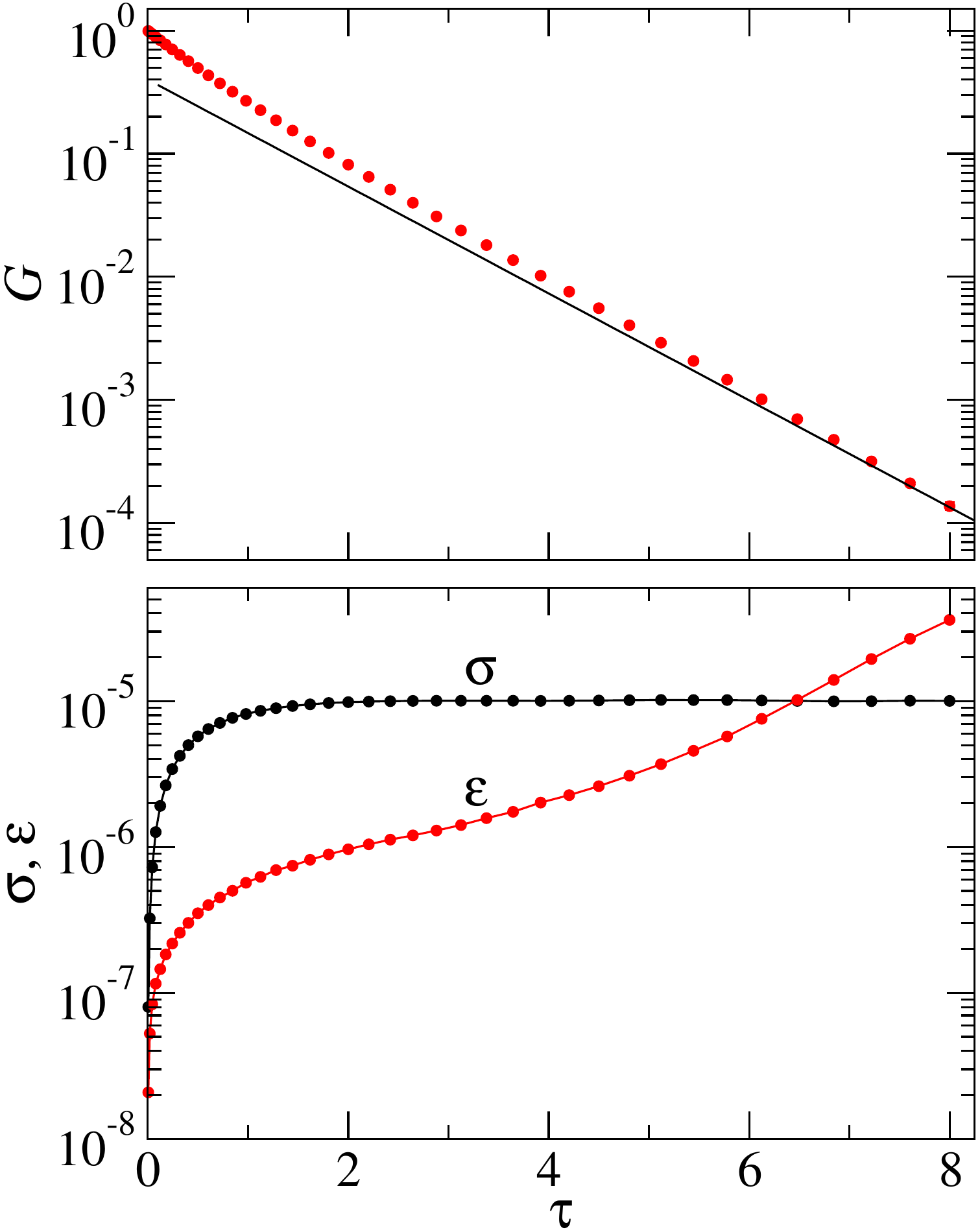}
\caption{The same kind of data as in Fig.~\ref{greal} but obtained using a synthetic spectrum with a $\delta$-function
of weight $a_0=0.4$ at $\omega_0=1$ and a continuum consisting of a half-Gaussian above the $\delta$-function, of width $1$.
The straight line in (a) corresponds to the contribution from just the $\delta$-function.}
\label{gsynt}
\end{figure}

In order to be able to test all aspects of the SAC procedures used with real QMC data, we generate a number $N_B$ of bins of noisy
data starting from the exact $G(\tau)$ computed from Eq.~(\ref{gtau}) with the given synthetic spectrum $S(\omega)$. These bins are
used to compute the mean values $\bar G_i$ and the covariance matrix with the same program used to process the QMC data. To construct
correlated noise similar to that present in QMC data, for each bin we first generate a set of normal-distributed random numbers $\sigma^0_i$,
with a given standard deviation (the same for all $i$, which is not necessarily exactly the case with QMC data but should be good enough
for testing purposes). We then run these data through a correlation procedure where a new noise set is generated according to
\begin{equation}
\sigma_i = \frac{\sum_{j} \sigma^0_j {\rm e}^{-|\tau_i-\tau_j|/\xi_\tau}}{\sqrt{\sum_{j}{\rm e}^{-2|\tau_i-\tau_j|/\xi_\tau}}}
\label{correlnoise}
\end{equation}
with a given autocorrelation time $\xi_\tau$. These noise values are then added to $G_i$. The autocorrelation time and the
original noise level $\sigma^0_i$ can be adjusted so that the eigenvalues of the covariance matrix are similar to those of typical
QMC data, though the QMC correlations can of course not be expected to exactly follow what is produced by Eq.~(\ref{correlnoise}).
An example is shown in Fig.~\ref{gsynt}, where we have adjusted the parameters of Eq.~(\ref{correlnoise}) to match the real QMC
data in Fig.~\ref{greal} closely (apart from an overall factor $\approx 2$ in the $\tau$-scale). As is apparent, we can indeed obtain
very similar forms of the standard errors and the eigenvalues of the covariance matrix.

\begin{thebibliography}{99}
\bibitem{Powalski17note}
The weight of the magnon pole was not reported in Ref.~\cite{Powalski17}, but the authors have confirmed to us that the
calculations do give weights similar to those in Refs.~\cite{Singh95} and \cite{sandvik01}.
\bibitem{formfactor}
In principle the cross-section also involves a form factor, corresponding to the finite extent of the orbitals hosting the localized spins,
but in most cases of interest the momentum-dependence of the form factor is very weak and can be neglected.
\end{thebibliography}
\end{document}